\documentclass[twocolumn,tighten]{aastex631}
 
\usepackage{amsmath}

\newcommand{\halpha}{H$\alpha$}
\newcommand{\hbeta}{H$\beta$}

\newcommand{\civ}{\ion{C}{4}}
\newcommand{\htwo}{H$_2$}

\newcommand{\av}{A$_{\text{V}}$}
\newcommand{\mdot}{$\dot{\text{M}}$}
\newcommand{\Mdot}{{\rm \dot{{M}}}}
\newcommand{\vsini}{$v\sin i$}

\newcommand{\msun}{\rm M_{\sun}}
\newcommand{\rsun}{\rm R_{\sun}}
\newcommand{\mstar}{\rm M_{\star}}
\newcommand{\rstar}{\rm R_{\star}}

\newcommand{\msunyr}{\rm{M_{\sun} \, yr^{-1}}}

\newcommand{\kms}{\rm \, km \, s^{-1}}
\newcommand{\escm}{\rm \, erg \, s^{-1} \, cm^{-2}}
\newcommand{\ri}{R$_{\rm i}$}
\newcommand{\rw}{W$_{\rm r}$}
\newcommand{\tmax}{T$_{\rm max}$}

\newcommand{\curf}{{\cal F}}

\newcommand{\pmm}[3]{#1^{+#2}_{-#3}}

\received{2024 June 21}
\revised{2024 September 13}
\accepted{2024 September 13}
\submitjournal{ApJ}
 
\shorttitle{C IV Model}
\shortauthors{Thanathibodee et al.}
 
\begin{document}
 
\title{A Model of the \ion{C}{4} $\lambda\lambda$ 1548, 1550 Doublet Line in T Tauri Stars}
 
\correspondingauthor{Thanawuth Thanathibodee}
\email{thanathi@bu.edu}
 
\author[0000-0003-4507-1710]{Thanawuth Thanathibodee}
\affiliation{Institute for Astrophysical Research, Department of Astronomy, Boston University, 725 Commonwealth Ave., Boston, MA 02215, USA}
 
\author[0000-0003-1639-510X]{Connor Robinson}
\affiliation{Department of Physics \& Astronomy, Amherst College, C025 Science Center 25 East Drive, Amherst, MA 01002, USA}
 
\author[0000-0002-3950-5386]{Nuria Calvet}
\affiliation{Department of Astronomy, University of Michigan, 1085 South University Ave., Ann Arbor, MI 48109, USA}
 
\author[0000-0001-9227-5949]{Catherine Espaillat}
\affiliation{Institute for Astrophysical Research, Department of Astronomy, Boston University, 725 Commonwealth Ave., Boston, MA 02215, USA}
 
\author[0000-0001-9301-6252]{Caeley Pittman}
\affiliation{Institute for Astrophysical Research, Department of Astronomy, Boston University, 725 Commonwealth Ave., Boston, MA 02215, USA}
 
\author[0000-0003-2631-5265]{Nicole Arulanantham}
\affiliation{Space Telescope Science Institute, 3700 San Martin Drive, Baltimore, MD 21218, USA}
 
\author[0000-0002-1002-3674]{Kevin France}
\affiliation{Laboratory for Atmospheric and Space Physics, University of Colorado Boulder, Boulder, CO 80303, USA}
 
\author[0000-0003-4243-2840]{Hans Moritz G\"unther}
\affiliation{Kavli Institute for Astrophysics and Space Research, Massachusetts Institute of Technology, 77 Massachusetts Ave., Cambridge, MA 02139, USA}
 
\author[0000-0002-0112-5900]{Seok-Jun Chang}
\affiliation{Max-Planck-Institut f\"{u}r Astrophysik, Karl-Schwarzschild-Stra$\beta$e 1, 85748 Garching b. M\"{u}nchen, Germany}
 
\author[0000-0002-5094-2245]{P. Christian Schneider}
\affiliation{Hamburger Sternwarte, Gojenbergsweg 112, D-21029, Hamburg, Germany}

\begin{abstract}
The C~{\sc iv} doublet in the UV has long been associated with accretion in T Tauri stars. However, it is still unclear where and how the lines are formed. Here, we present a new C~{\sc iv} line model based on the currently available accretion shock and accretion flow models. We assume axisymmetric, dipolar accretion flows with different energy fluxes and calculate the properties of the accretion shock. We use Cloudy to obtain the carbon level populations and calculate the emerging line profiles assuming a plane-parallel geometry near the shock. Our model generally reproduces the intensities and shapes of the C~{\sc iv} emission lines observed from T Tauri stars. We find that the narrow component is optically thin and originates in the postshock, while the broad component is optically thick and emerges from the preshock. We apply our model to seven T Tauri stars from the Hubble Ultraviolet Legacy Library of Young Stars as Essential Standards Director's Discretionary program (ULLYSES), for which consistently determined accretion shock properties are available. We can reproduce the observations of four stars, finding that the accretion flows are carbon-depleted. We also find that the chromospheric emission accounts for less than 10 percent of the observed C~{\sc iv} line flux in accreting T Tauri stars. This work paves the way toward a better understanding of hot line formation and provides a potential probe of abundances in the inner disk.
\end{abstract}
 
 
\defcitealias{calvet1998}{CG98}
 
\section{Introduction} \label{sec:intro}
Magnetospheric accretion has been the paradigm of accretion in low-mass, pre-main sequence stars for more than two decades, with support from theoretical and observational studies.  Numerical models of magnetospheric accretion in young pre-main sequence stars (T Tauri stars; TTS) have been developed to explain observations in hydrogen lines to great length \citep{hartmann1994,muzerolle1998a,muzerolle2001,bouvier2007,kurosawa2013,esau2014,tessore2021}. Applying these models has provided insight into the overall structure of the magnetospheric flow, barring some assumptions. Similarly, advances in infrared interferometric observations of the closest T Tauri stars \citep[e.g.,][]{gravity-collaboration2020,gravity-collaboration2023ix,gravity-collaboration2023x}  have confirmed this picture of accretion onto young stars.
 
An accretion shock on the stellar surface is one of the consequences of magnetospheric accretion in T Tauri stars \citep[see][for review]{hartmann2016}. A shock is formed as the accretion flow material reaches the photosphere near free-fall velocities (hundreds of $\kms$). The kinetic energy from the flow is reprocessed and heats the region before the shock (the preshock), immediately after the shock (the postshock), the underlying photosphere (the heated photosphere), as well as its immediate surroundings \citep{calvet1998,colombo2019}. Models of the accretion shock have been used to explain the excess continuum emission in the near-UV of accreting TTS \citep{calvet1998}. Variations of such models have been used to infer the structure of the accretion shock hotspots \citep{ingleby2013,espaillat2021,robinson2022,pittman2022}, suggesting the density stratification at the base of the accretion flow. Such stratification is also inferred from X-ray observations \citep[e.g.,][]{schneider2018oct}.
 
In addition to continuum excess emission, the regions associated with the accretion shock are also thought to produce emission lines. However, where and how observed lines form is still uncertain. Much progress has been made in using X-ray emission lines to improve our understanding of properties of accretion shock regions, such as densities, temperature, geometry, and chemical abundances \citep[e.g.,][]{kastner2002,stelzer2004,drake2005,argiroffi2005,brickhouse2010,argiroffi2017}. Shock-originated emission lines are also found in the optical, such as \ion{He}{1}\,$\lambda5876$, which is used to determine magnetic obliquity in TTS \citep{mcginnis2020}.

Atomic emission lines originating in the magnetospheric flow from the inner disk onto the star also provide an indirect probe of the inner disk composition \citep{micolta2023}. Carbon is an especially critical tracer of this region, as it is one of the most abundant elements in the Universe after hydrogen and helium. It is a key element of (Earth-like) life, and carbon-bearing species play important roles in the chemistry of protoplanetary disks \citep[e.g.,][]{oberg2023}. Therefore, detecting and characterizing carbon-bearing species can provide many insights into the properties of protoplanetary disks, including their chemical composition, density, and temperature structures. Carbon-bearing molecules have been studied extensively in the sub-mm and infrared using facilities such as ALMA, the VLT, and JWST. These lines form in the colder regions of protoplanetary disks. Probing the inner disk is more challenging. The \ion{C}{1} atomic lines in the near-infrared (e.g., 0.908, 0.94, 1.069, 1.176, 1.45, 2.166 $\mu$m) provide a way, but they have been observed in only a handful of sources \citep{mcclure2019,mcclure2020}.
 
Another way to indirectly probe the inner disk is by studying the magnetospheric flow from the disk onto the star. Since the flow's temperatures are high, all elements are expected to be atomic and possibly ionized. Therefore, studying atomic emission lines formed in the accretion flow or the shock can be used to determine the elemental compositions of the inner disk \citep{micolta2023}. 
 
The {\civ} lines at 1548 and 1550\,{\AA} are some of the most prominent lines in the spectra of T Tauri stars (see review by \citealt{schneider2020}), and they were recognized very early to be related to accretion \citep{ardila2002,johns-krull2000,valenti2000,calvet1996,valenti1993b}. Each of the lines in the doublet is generally composed of a narrow component (NC) with low redshifted velocity (a few $10\,\kms$) and a broad component (BC) with velocity widths on the order of a few hundred $\kms$ and a wide range of the centroid velocity ($\sim-20$ to $\sim+200\,\kms$; \citealt{ardila2013}). Using extensive data from the \textit{HST} DAO program (PI: G. Herczeg, PID: 11616), \citet{ardila2013} analyzed the {\civ} lines of 28 TTSs and confirmed previous studies that the line luminosity correlates with mass accretion rates, suggesting that the line forms in the accretion process. The line profiles, on the other hand, are harder to explain. The Gaussian decomposition results of \citet{ardila2013} suggest that only half of the CTTS in their sample are consistent with the scenario where the BC is formed in the preshock, as the centroid velocity of the BC is less than four times that of the NC, as predicted by the jump condition for the shock. However, as we will discuss later, we note that their interpretation does not consider geometry effects, such that the emission from the preshock may peak at lower velocities.

Several theoretical works have attempted to explain the observed line profiles to study how the line forms. \citet{lamzin2003a,lamzin2003b} calculated the properties of the accretion shock and the {\civ} lines using axially symmetric shock and radial accretion and found that the model profiles are inconsistent with the observations. \citet{colombo2016} synthesized the {\civ} line from MHD simulations with fragmented accretion streams and trains of accretion blobs and found a reasonable agreement between the synthetic line profiles and the observed profiles, but the line flux levels are one to two orders of magnitude lower than that of the observations. In this work, we expand upon the work of \citet{lamzin2003b} and present a new model of the {\civ} line formation based on the accretion shock model of \citet{calvet1998} and the accretion flow model of \citet{muzerolle2001}. The goals of this paper are twofold: (1) to understand where and how the {\civ} line forms in the context of the accretion shock and (2) to use the {\civ} line to infer properties of the innermost disk regions around accreting T Tauri stars. We describe our model in Section~\ref{sec:model} and provide some model results in Section~\ref{sec:model_results}. In Section~\ref{sec:orion}, we apply the model to a subset of ULLYSES targets in Orion OB1 and discuss the results in Section~\ref{sec:discussion}. We provide conclusions and a summary in Section~\ref{sec:summary}.

\begin{figure*}[t]
\epsscale{1.15}
\plotone{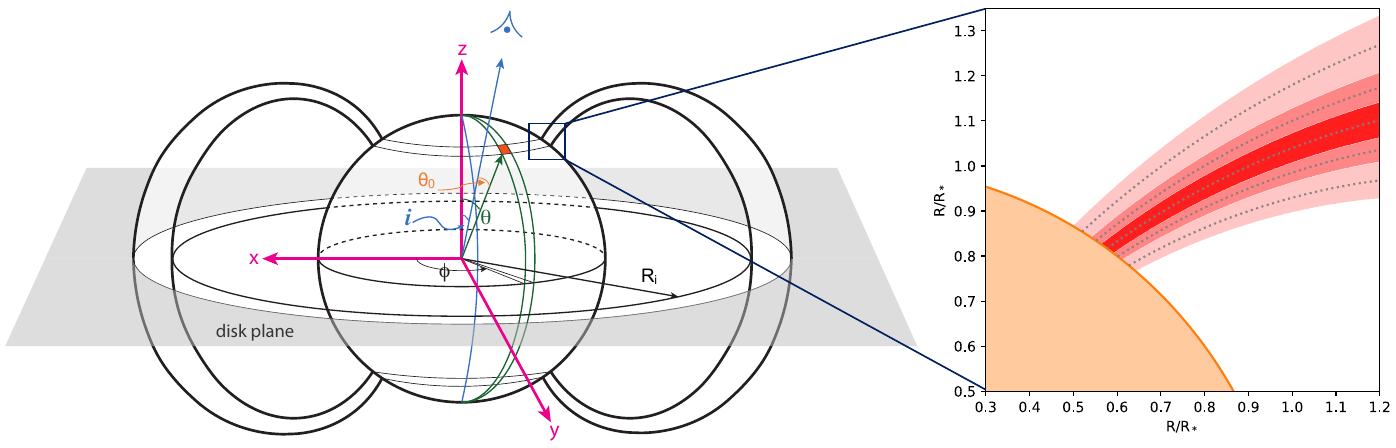}
\caption{\emph{Left:} The overall geometry of the model. The star's equatorial plane is the same as that of the disk. Axisymmetric accretion funnels originate from the disk and land perpendicular to the star's surface. The vector toward the observer is defined to be in the $yz$ plane with the colatitude $i$, i.e., the inclination of the star.  \emph{Right:} A zoom-in, side-view illustration of the large-scale geometry of accretion flows landing on the surface of a star. The darker color corresponds to the highest energy flux ($\curf_h$). The post/preshock are small regions at the base of the flows. Gray dotted lines are the center of the flow used to define the colatitude $\theta$ for a given flow.
\label{fig:geometry}}
\end{figure*}
 
\section{Description of the Model} \label{sec:model}
As in previous studies \citep{ardila2013,lamzin2003a,lamzin2003b}, we consider that the {\civ} doublet forms near the shock created by accreting material reaching the photosphere. In this section, we provide a detailed description of the model and the line calculation.

\subsection{Geometry}
The large-scale geometry of the model is illustrated in Figure~\ref{fig:geometry}. A given point on the star's surface is defined using spherical coordinates ($\theta$, $\phi$) centered at the star. $\theta$ is measured from the pole, and $\phi$ is defined such that the vector pointing toward the Earth is ($\theta$, $\phi$)=($i$, $\pi/2$), where $i$ is the star's inclination. We assume that the shock is created at the base of the axisymmetric accretion flow in a dipole geometry, which lands perpendicular to the surface. As a result, the shock surface is an annulus on the star. Within this framework, a filling factor $f'$ subtends an area on the stellar surface between any two co-latitudes $\theta_1$ and $\theta_2$ as
\begin{equation}
    f'=\frac{1}{2}\left[\cos(\theta_1) - \cos(\theta_2)\right].
\end{equation}
 
Similarly to \citet{pittman2022}, we assume that the ring is composed of multiple flows at three different energy fluxes: high energy flux $\curf_h$, medium energy flux $\curf_m$, and low energy flux $\curf_l$. Motivated by results from numerical simulations and observations \citep[e.g.,][]{romanova2004a,espaillat2021}, we assume that two medium-energy flux flows blanket the central flow, which has the highest energy flux (and highest density), and together they are sandwiched by two low-energy flux flows. Therefore, there are five rings in total for each hemisphere (c.f., Figure~\ref{fig:geometry}). The same geometry also occurs in the other hemisphere since we expect that the material in the accretion flows lands on both hemispheres. That is
\begin{equation}
    f_{h,m,l} = 2f'_{h,m,l},
\end{equation}
where $f$ is the total filling factor of a given energy flux.
 
The co-latitude for each ring is defined by two associated magnetic radii $R_{in}$ and $R_{out}$  through the dipolar flow equation as
\begin{align}
\theta &= \frac{1}{2}\left(\theta_{in}+\theta_{out}\right) \\
        &= \frac{1}{2}\left[\sin^{-1}\left(\sqrt{\frac{R_*}{R_{in}}}\right) + \sin^{-1}\left(\sqrt{\frac{R_*}{R_{out}}}\right)\right].
\end{align}
Figure~\ref{fig:geometry} shows the large-scale geometry of the multi-component flow. We assume that the flows at different energy fluxes are independent. We also assume that the surrounding, large-scale accretion flows ($\sim$few stellar radii) do not absorb any emission emerging from the shock.

As in \citet{calvet1998}, the shock at the star's surface is composed of the preshock and the postshock region. For simplicity, we assume that the flow lands perpendicular to the star's surface and that the shock structure's height ($\sim$ the shock cooling distance) is much smaller than the stellar radius. Therefore, we assume plane-parallel geometry for the following radiative transfer calculation.

For a given spot $(\theta, \phi)$ on the surface, it can be shown \citep[e.g.,][]{gunther2011} that the angle between the line of sight and the flow is given by
\begin{equation}
\cos\theta_0 = \sin\theta\sin\phi\sin(i) + \cos\theta\cos(i), \label{eq:theta0_up}
\end{equation}
for the upper hemisphere ($\theta<\pi/2$), and
\begin{equation}
\cos\theta_0 = \sin\theta\sin\phi\sin(i) - \cos\theta\cos(i), \label{eq:theta0_low}
\end{equation}
for the lower hemisphere ($\theta>\pi/2$). We only include in our calculation the point ($\theta$, $\phi$) where $\theta_0<\pi/2$, since the flow would be behind the star for a larger angle. This requirement limits the longitude for the ring to be seen between $\phi_w$ and $\phi_e$.
 
\subsection{Shock Structure and Atomic Level Population} \label{ssec:shock_struct}
\subsubsection{Initial Calculations} \label{sssec:struct_init}
\begin{figure}[t]
\epsscale{1.0}
\plotone{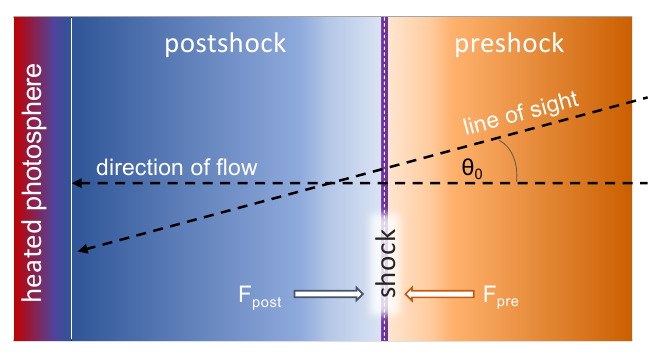}
\caption{The schematic view of an accretion column and its associated geometry. An observer is viewing the flow at an angle $\theta_0$.
\label{fig:geometry_shock}}
\end{figure}

We use the 1D model of \citet{calvet1998}, as updated by \citet{robinson2019}, to calculate the structure of the preshock and postshock, and we refer the reader to those papers for details. The postshock density and temperature are calculated by solving the fluid equations coupled with hydrostatic equilibrium equations using emissivities from Cloudy \citep{ferland2017}. The resulting emergent spectrum from the postshock $F_{post}$ is then used as an input to calculate the emergent spectrum of the preshock $F_{pre}$, using Cloudy. These spectra are used, without modification, for subsequent runs to calculate ionization, continuum opacities, and carbon level populations. Figure~\ref{fig:geometry_shock} shows the geometry of the shock and the level population calculation.

\subsubsection{Subsequent Calculations}
After the initial run of the model to generate $F_{post}$ and $F_{pre}$, we use Cloudy to calculate the relevant properties of the post and preshock for line radiative transfer. Cloudy is generally used to determine the steady-state properties of a slab of material irradiated by an external source given a density structure and, optionally, a temperature structure. In our model, the postshock is irradiated by the emission from the preshock ($F_{pre}$); similarly, the preshock is irradiated by the postshock ($F_{post}$). Some previously calculated properties of the slabs are fixed. The density and temperature structure of the postshock are fixed from the shock model calculation. For the preshock, we assume a constant density; Cloudy calculates the temperature along with other properties. In both the postshock and the preshock, we also include turbulent velocities $v_{t, pre/post}$ of the flow and carbon abundance in the solar scale $[C/H]=\log(N_{C}/N_{H})_{\star} - \log(N_{C}/N_{H})_{\odot}$, as parameters. In these Cloudy runs, we record the level population and other properties as a function of depth from the shock.
 
\subsection{Line Profile Calculation}
Since the post and preshock heights are smaller than the stellar radius, we assume plane-parallel geometry to calculate the line profiles. The overall flux is given by,
\begin{equation}
    F_{\nu} = \frac{R_*^2}{d^2}\int_{\phi_E}^{\phi_W}\left(I_{\nu,shock}+I_{\nu,\star}\right)\cos\theta_0\sin\theta \Delta\theta d\phi, \label{eq:fnu}
\end{equation}
where we assume that the ring is thin and $\Delta\theta=(\theta_{in}-\theta_{out})$, $R_*$ is the radius of the star, and $d$ is its distance. $\phi_E$ and $\phi_W$ refer to the easternmost and westernmost edges of the ring, at which point $\theta_0=\pi/2$. We assume a blackbody at the star's effective temperature for $I_{\nu, star}$. The $\theta_0$-dependent $I_{\nu, shock}$ is composed of the contributions from the preshock and the postshock. We assume that the postshock is always viewed through the preshock, therefore
\begin{equation}
    I_{\nu, shock} = I_{\nu, pre} + I_{\nu, post}e^{-\tau_{\nu, tot, pre}},
\end{equation}
where $\tau_{\nu, tot, pre}$ is the total optical depth of the preshock. We follow the same steps to calculate the specific intensity and the optical depth of the preshock and the postshock, as shown below.
 
For a given flow, the specific intensity at a position ($\theta$, $\phi$) depends on the angle between the accretion flow direction and the line of sight $\theta_0$ (c.f., Figure~\ref{fig:geometry}). The plane-parallel assumption simplifies its calculation to the integration of the emissivity along the flow toward the star along $r$ with a factor $\mu$ as
\begin{align}
    I_{\nu}(\mu) &= \int_{r_0}^{\infty}(\eta_l(r) + \eta_c(r))e^{-\mu\tau_{\nu}(r)}\mu dr, \label{eq:inu}\\
\mu &\equiv \frac{1}{\cos\theta_0},
\end{align}
where $\eta_l$ and $\eta_c$ are the line and continuum emissivity, respectively, and $\tau$ is the optical depth.
 
The continuum emissivity is taken directly from Cloudy simulations. The line emissivity is 
\begin{equation}
\eta_{\nu, l} = \chi_{\nu, l}S_l,
\end{equation}
 where the line opacity is given by
\begin{align}
\chi_{\nu, l} &= \chi_{0, l}\cdot V_{\nu}(\theta_0, v_{t}, v(r))\\
        &=\frac{h\nu_0}{4\pi}B_{lu}N_l\left(1-\frac{N_ug_l}{N_lg_u}\right)\cdot V_{\nu}(\theta_0, v_{t}, v(r)),
\end{align}
 
where
\begin{equation}
    B_{lu} = \frac{g_u}{g_l}B_{ul} = \frac{g_u}{g_l}\frac{c^2}{2h\nu_0^3}A_{ul},
\end{equation}
 
and line source function is given by
 
\begin{equation}
    S_l = \frac{2h\nu_0^3}{c^2}\left[ \frac{N_lg_u}{N_ug_l} -1 \right]^{-1}. \label{eq:source_function}
\end{equation}
 
The Voigt profile $V_{\nu}$ depends on the turbulent velocity $v_t$, assumed to be a constant parameter for a given flow, and the local properties, including the velocity $v(r)$. It is calculated as
\begin{equation}
    V_{\nu} = \frac{1}{\pi\Delta\nu_D}H(\frac{\Gamma_r}{4\pi\Delta\nu_D}, \frac{\nu-\nu'}{\Delta\nu_D}),
\end{equation}
where $H$ is the Hjerting function and $\Gamma_r$ is the line radiative broadening parameter. Stark broadening is small for the electron densities considered here ($\sim10^{12}-10^{15}$; \citealt{elabidi2011}), while the van der Waals broadening is expected to be minimal in the ionized environment that the line forms. The Doppler width is calculated from the thermal and turbulent velocities, $\Delta\nu_D = \frac{\nu_0}{c}\sqrt{v_{term}^2 + v_{t}^2}$, and the shifted line-center frequency is $\nu' = \nu_0(1-v(r)\cos\theta_0/c)$. In the preshock, $v(r)$ is a constant, adopted as a free-fall velocity from $R_{in}$; $v_{pre}=\sqrt{2GM_*(1/R_*-1/R_{in})}$ \citep{calvet1998}. In the postshock, $v(r)$ decreases from $\frac{1}{4}v_{pre}(r)$ at the shock to $\sim0$ at the photosphere. $v_{post}(r)$ is obtained along with the postshock structure in Section~\ref{sssec:struct_init}. 
 
The optical depth along the flow is given by
\begin{equation}
\tau_{\nu}(r) = \int_{r}^{\infty}
\left (\chi_{\nu,l}(r) + \chi_c(r)
\right )dr,
\end{equation}
where the continuum opacity $\chi_c$ is computed by Cloudy.
 
All relevant atomic parameters are retrieved from NIST \citep{NIST_ASD} and are summarized in Table~\ref{tab:atomic_param}.
 
\subsection{Model Limitations} \label{ssec:limitations}
We make several approximations and assumptions to our model, and here we list these limitations.  These assumptions are made while keeping the following goal in mind: to study how the geometry of the emitting region and the large-scale accretion properties affect the {\civ} line.
 
First, several works have shown that the accretion shock is dynamic, and time-dependent models may be relevant \citep[e.g.,][]{de-sa2019}. We note, however, that the timescale of quasi-periodic oscillation (QPO) cycles in the time-dependent models are on the order of 100-600\,s. In contrast, the observational timescales (e.g., telescope exposure times) are on the order of 1000-3000\,s or more. Dedicated analysis of TW Hya's observations with sufficient time resolution also found no evidence of this variability (\citealt{drake2009,gunther2010}, but, see also \citealt{siwak2018}). Other MHD simulations of accretion shocks, such as those presented in \citep{orlando2010}, show that these oscillations disappear in the 2D analysis. Therefore, a time-average model will be needed to explain the observation, and we opt to use steady-state approximation following most of the literature on this topic.
 
Second, we used a readily available and widely utilized code, Cloudy, to calculate the population. This code uses the \emph{escape probability method} for radiative transfer, and the results are not exact \citep{hubeny2001}. According to the Cloudy documentation \emph{Hazy 2}, this approximation may be problematic for transitions where the lower and upper levels are in the excited states, but it notes that resonance lines (e.g., {\civ}, with the lower level at electronic ground state) should be fine. Therefore, as a first approximation, we adopted Cloudy's formalism.
 
Third, we adopted a strictly 1D, plane-parallel geometry in calculating the flow structure and level population. That is, we do not consider the angular dependence of the irradiation in the steps described in Section~\ref{ssec:shock_struct}, following \citet{calvet1998} and \citet{robinson2019}. This omission may introduce some uncertainty in the level population and, subsequently, the line flux, especially near the edge of the flow. However, as pointed out by observations and simulations \citep[e.g.,][]{espaillat2021}, the outer parts of the flow may have a lower density than the inner part, and this part contributes less to the overall flux/profile, as will be shown later in this work. Therefore, the 1D approximation is justified.

\begin{deluxetable}{lccc}
\tablecaption{Atomic Parameters of the {\civ} lines.
\label{tab:atomic_param}}
\tablehead{
\colhead{Parameter} &
\colhead{{\civ} $\lambda$1548} &
\colhead{{\civ} $\lambda$1550} &
\colhead{Unit}
}
\startdata
$\lambda_{\mathrm{obs}}$ & 1548.202 & 1550.774 & \AA \\
Level Energy & & & \\
\hspace{0.3cm} $\mathrm{E_l}$ & 0.0 & 0.0 & eV \\
\hspace{0.3cm} $\mathrm{E_u}$ & 8.00835 & 7.99500 & eV \\
Config. Terms & & & \\
\hspace{0.3cm} Lower & 1s$^2$2s $^2$S$_{1/2}$ &  1s$^2$2s $^2$S$_{1/2}$ & \\
\hspace{0.3cm} Upper & 1s$^2$2p $^2$P$_{3/2}$ & 1s$^2$2p $^2$P$_{1/2}$ & \\
Stat. weight $g_l \rightarrow g_u$ & $2\rightarrow4$ & $2\rightarrow2$ & \\
Transition Prob. $\mathrm{A_{ul}}$ & $2.65\times10^8$ & $2.64\times10^8$ & s$^{-1}$ \\
Rad. broadening $\Gamma_r$ & $2.65\times10^8$ & $2.64\times10^8$ & s$^{-1}$ \\
Oscillator strength $f_{lu}$ & 0.19 & 0.0952 & 
\enddata
\end{deluxetable}

\begin{figure*}[ht]
\epsscale{0.9}
\plotone{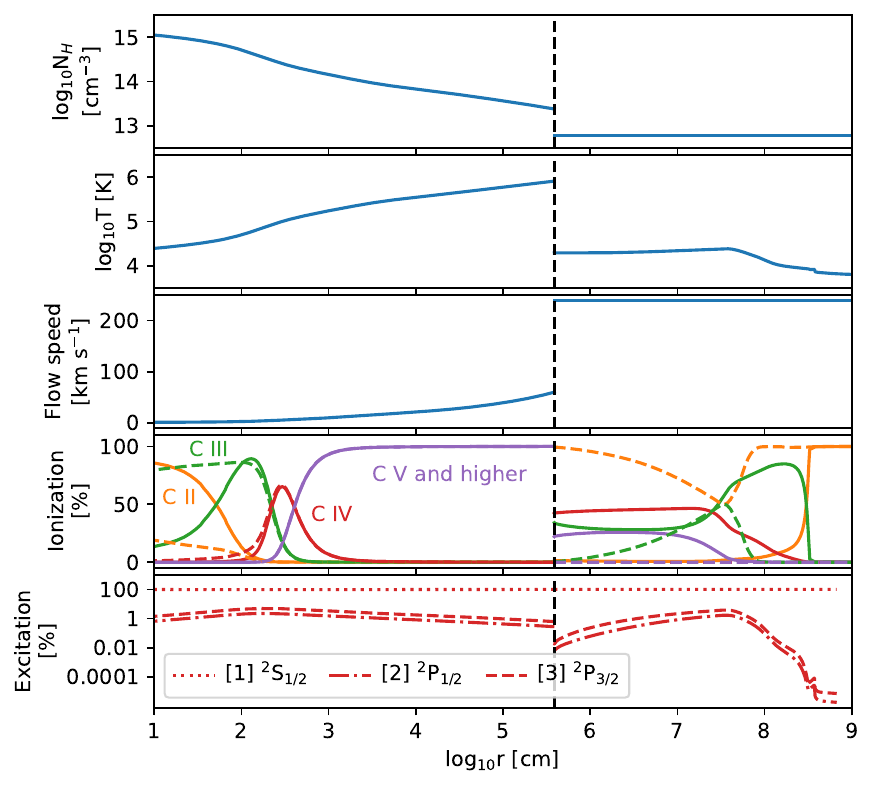}
\caption{The macroscopic and microscopic properties of an accretion flow with $\curf=10^{11}\escm$, $\mstar=0.37\,\msun$, and $\rstar=1.66\,\rsun$. The horizontal axis indicates the distance from the transition between the heated photosphere and the postshock. The vertical dashed line marks the location of the shock. In the fourth row, the dashed lines indicate the fraction at each ionization state, assuming LTE, for comparison.
\label{fig:flow_properties}}
\end{figure*}
 
\section{Model Results} \label{sec:model_results}
In this section, we showcase the results using the models calculated for CVSO~104 ($\mstar=0.37\,\msun$, $\rstar=1.66\,\rsun$; \citealt{manara2021}) and explore various properties of the flows. We also study the effects of varying turbulent velocities ($v_t=10$ to $100\,\kms$), carbon abundances (${\rm [C/H]}=0$ to $-3$), inclinations ($i=0\degr$ to $90\degr$), and truncation radii ($3.14$ and $5$\,{$\rstar$}) on the {\civ} line properties and profiles.

\subsection{Properties of the flows} \label{ssec:flow_properties}

After solving the density and velocity structure of the postshock and preshock and the temperature of the postshock (Section~\ref{ssec:shock_struct}), we calculate the atomic properties related to the line profile calculation (see Section~\ref{ssec:shock_struct}.) Figure~\ref{fig:flow_properties} shows the structure of the flow and the resulting ionization and level population. As in Figure~\ref{fig:geometry_shock}, the observer is viewing the flow from the right through the preshock, toward the shock, and finally, the postshock.

By construction, the hydrogen number density  $N_H$ and the flow velocity $v(r)$ of the preshock are constant as the depth of the preshock is much smaller than the star's radius. After the shock, the temperature and density of the gas sharply increase while the speed suddenly drops, as expected. The fourth row of Figure~\ref{fig:flow_properties} shows the ionization fraction of the carbon atom. In the postshock, {\civ} is only present in a thin, $\sim10^3$\,cm layer near the transition to the heated photosphere.  In contrast, {\civ} is the dominant ionization level for a much thicker ($\sim10^7$\,cm) region in the preshock.

\begin{figure*}[t]
\epsscale{0.9}
\plotone{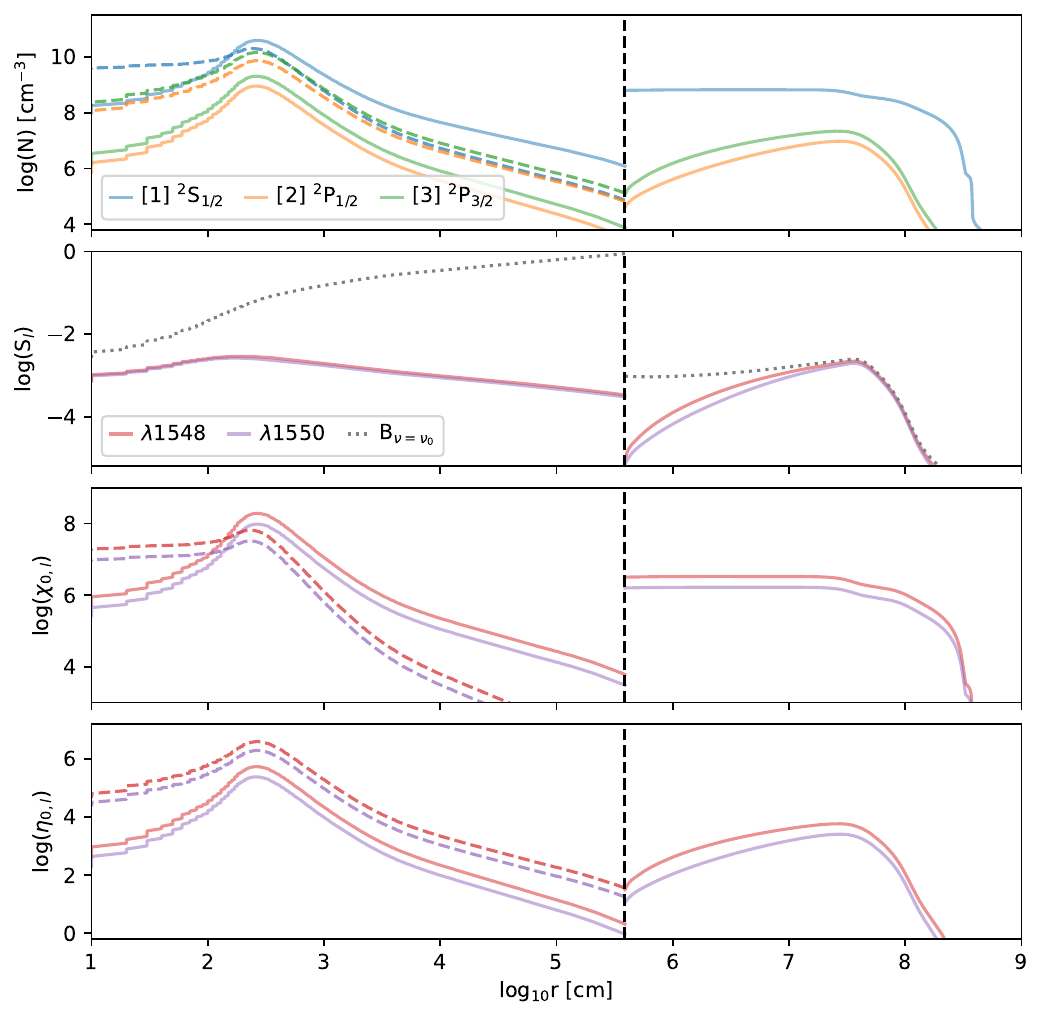}
\caption{Atomic and radiative properties of the postshock and preshock flows with $\curf=10^{11}\escm$ and $\mstar=0.37\,\msun$, $\rstar=1.66\,\rsun$. The shock is marked with the vertical dashed line, and the distance is measured outwardly from the postshock-heated photosphere transition. The observer is viewing through the flow from the right toward the star on the left. In the postshock, the dashed lines show the quantities with the LTE level populations. The LTE populations in the preshock are much lower than the range presented here and are not shown. In the second row, the dotted grey line is the Planck function, plotted here for comparison. The Planck function for the postshock traces the same curve as the LTE source function (dashed line). The units for $S_l$, $\chi_{0,l}$, and $\eta_{0,l}$ are ${\rm erg\,s^{-1}\,cm^{-2}\,sr^{-1}\,Hz^{-1}}$, ${\rm cm^{-1}}$, and ${\rm erg\,s^{-1}\,cm^{-3}\,sr^{-1}\,Hz^{-1}}$, respectively.
\label{fig:line_properties}}
\end{figure*}
 
The {\civ} doublet lines are formed between the transitions of the first three electronic levels of C$^{3+}$: [1] $^2$S$_{1/2}$, [2] $^2$P$_{1/2}$, and [3] $^2$P$_{3/2}$ (c.f., Table~\ref{tab:atomic_param}). The last row of Figure~\ref{fig:flow_properties} shows the percentage of each electronic level of the {\civ} ionization level. In every region, most electrons in the {\civ} atoms are in the ground state, as the other levels account for $\sim1\%$ of the population. In the top row of Figure~\ref{fig:line_properties}, we compare the number density of the {\civ} electronic states across the shock region. While the density of the preshock is lower due to a higher velocity and constant mass flux, the number densities of {\civ} are comparable to the peak number density in the postshock and at a much larger extent. For comparison, we plot the LTE level population assuming the density and temperature of the given region, shown as dashed lines at the top of Figure~\ref{fig:line_properties}. The number densities in the postshock depart slightly from LTE, with the densities peaking at the same location at $\sim10^{2.5}\,$cm. Closer to the shock, the {\civ} number densities decrease quickly as the atoms are ionized due to the high temperatures. The LTE population is much lower in the preshock (not shown), suggesting that the population is dominated by photoionization and radiative excitation from the shock.

The second row of Figure~\ref{fig:line_properties} shows the line source function ($S_l$; Equation~\ref{eq:source_function}). The source function varies slowly with the distance in the postshock with only a small slope, whereas in the preshock, the source function peaks at a certain distance and drops off quickly closer/further from the shock. The Planck function is plotted as dotted lines in the second row for comparison. The condition differs from LTE throughout the postshock. On the other hand, the source function is essentially the Planck function at the outer end of the preshock, further away from the shock, as conditions approach LTE.
 
\begin{figure*}[t]
\epsscale{0.9}
\plotone{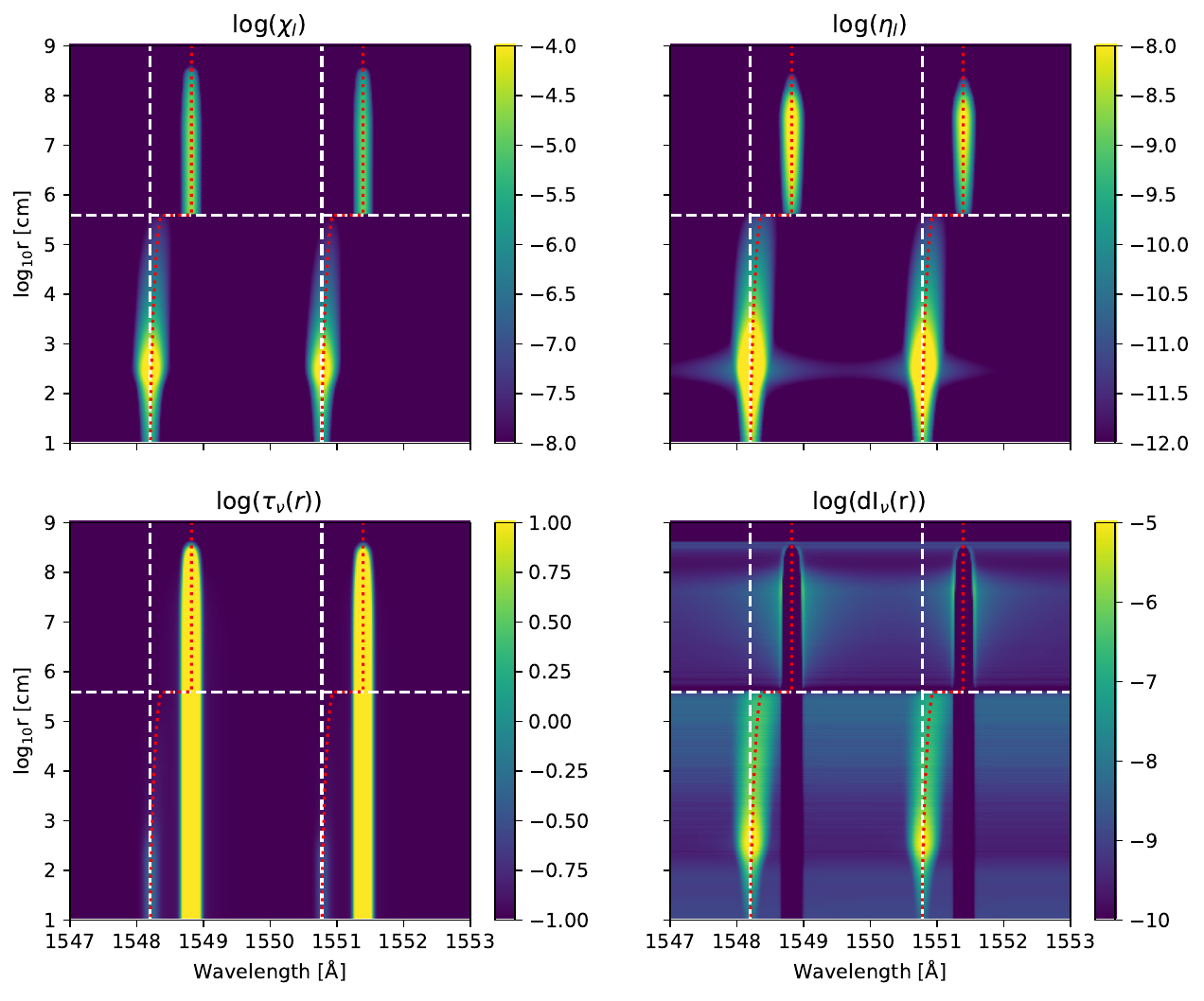}
\caption{The wavelength-dependent, two-dimensional line properties for the same model as in Fig.~\ref{fig:line_properties} for $\cos\theta_0=0.5$. The values in the upper row are the local properties of the lines, whereas the values in the lower row are cumulative and include the contribution from the continuum. The white dashed vertical lines mark the rest-frame line centers, and the horizontal lines mark the shock location. The red dotted lines show the shifted line-center wavelength $\lambda'=\lambda_0(v(r)\cos\theta_0/c+1)$.
\label{fig:line_2d}}
\end{figure*}
 
The line-center opacity ($\chi_{0,l}$) is shown in the third row of Figure~\ref{fig:line_properties}. Similar to the population (first row), $\chi_{0,l}$ peaks in a narrow region in the postshock, whereas it is more constant in the preshock until a fast drop off. In the postshock, the opacity is higher than in the LTE cases when the flow is closer to the shock and lower than in the LTE cases when approaching the heated photosphere. In the preshock, it is higher than LTE at all distances. The resulting line-center emissivity, $\eta_{0}=\chi_0S_l$, is shown in the last row of Figure~\ref{fig:line_properties}. The postshock has largely higher emissivity than the preshock. However, this does not translate to intensity.
 
In Figure~\ref{fig:line_2d}, we show the wavelength-dependent line properties of the same model as in Figure~\ref{fig:line_properties}, assuming $\mu\equiv1/\cos\theta_0=2$, but the results are similar regardless of angle. The top row of Figure~\ref{fig:line_2d} shows the opacity $\chi_{\nu, l}$ and the emissivity $\eta_{\nu, l}$ as a function of depth. The opacity and emissivity peak near the rest-frame line centers (vertical dashed lines) in the postshock, whereas a velocity shift is seen for the preshock, as expected. They are also broader and higher, suggesting that conditions in the postshock are more suitable for line formation than that in the preshock. Specifically, the higher temperatures mean that the lines are broader (higher thermal width), and the higher density means higher $\chi$ and $\eta$.
 
The bottom row of Figure~\ref{fig:line_2d} shows the optical depth from the observer toward the star along the flow (left panel) and the differential intensity from each depth (right panel). The continuum opacity and emissivity are included in these calculations. Even though the opacity in the preshock is lower than in the postshock, the line optical depth is much higher in the preshock than in the postshock due to a larger physical distance. The optical depth is low throughout the postshock near the rest-frame line centers.
 
The differential intensity $dI_{\nu}=(\eta_l+\eta_c)\,e^{-\mu\tau_{\nu}}\,\mu\,dr$ (c.f., Equation~\ref{eq:inu}) is shown in the lower right panel of Figure~\ref{fig:line_2d}.  Due to the high optical depth, the preshock emission near the shifted line center $\lambda'=\lambda_0(v(r)\cos\theta_0/c+1)$ (red dotted lines) is absorbed for nearly the entire flow. Only the radiation from the two wings of each line can escape the flow and contribute to the final line emission. On the other hand, the low optical depth in the postshock allows the radiation to escape easily. Again, we note that the differential intensity in the preshock is lower than in the postshock, but its sum is higher due to the much larger thickness.
 
Lastly, the Figure also shows that the emission from the preshock comes mostly from the front of the flow (i.e., closer to the observer), but most emission from the postshock comes from the back (i.e., the interface between the postshock and the heated photosphere beneath). This is where the temperature has decreased enough for {\civ} to be the dominant ionization stage, as shown in Figure~\ref{fig:flow_properties}. This is also where the velocity in the flow is lowest, which explains the smaller shift in velocity space from the narrow component of the line as seen in the observations \citep{ardila2013}.

\begin{figure*}[t]
\epsscale{0.9}
\plotone{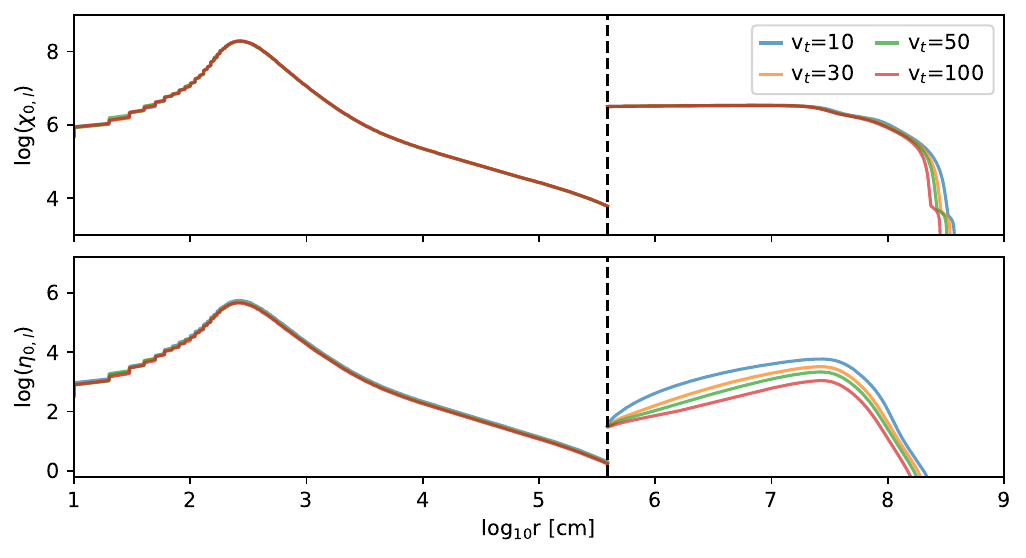}
\caption{The change in radiative properties of the flow due to turbulence velocities. The units are the same as in Fig.~\ref{fig:line_properties}.
\label{fig:diff_vt}}
\end{figure*}

\subsection{The Effects of Turbulence} \label{ssec:effects_turb}

Multidimensional simulations of accretion shocks have shown that the process can be very turbulent, depending on the ratio of gas/magnetic pressure. In the preshock, the magnetic pressure is much stronger than gas pressure, and the mass follows the field lines, so turbulence should not be dominant. In the shock and the postshock,  the kinetic energy of the bulk flow is converted into heat and/or turbulence. The hydrodynamic pressure overcomes the magnetic pressure, and turbulent flows can develop \citep{orlando2010}.

Nevertheless, the level of turbulence in the flow is unknown, so we model it as inherent turbulence velocity. We investigate this effect on the radiative properties of the flows by varying the velocity in the Cloudy input. Figure~\ref{fig:diff_vt} shows the line-center opacity and emissivity for the postshock and preshock. Changing the turbulent velocity does not affect the radiative properties of the postshock. In the preshock, the turbulence affects the emissivity in the entire flow and the opacity in the region where the line approaches LTE (c.f., Figure~\ref{fig:line_properties}). This is because turbulence affects the pressure and temperature, affecting the level population as it approaches LTE further away from the radiation front.

\subsection{The Effects of Abundance} \label{ssec:effects_abun}
\begin{figure*}[t!]
\epsscale{0.9}
\plotone{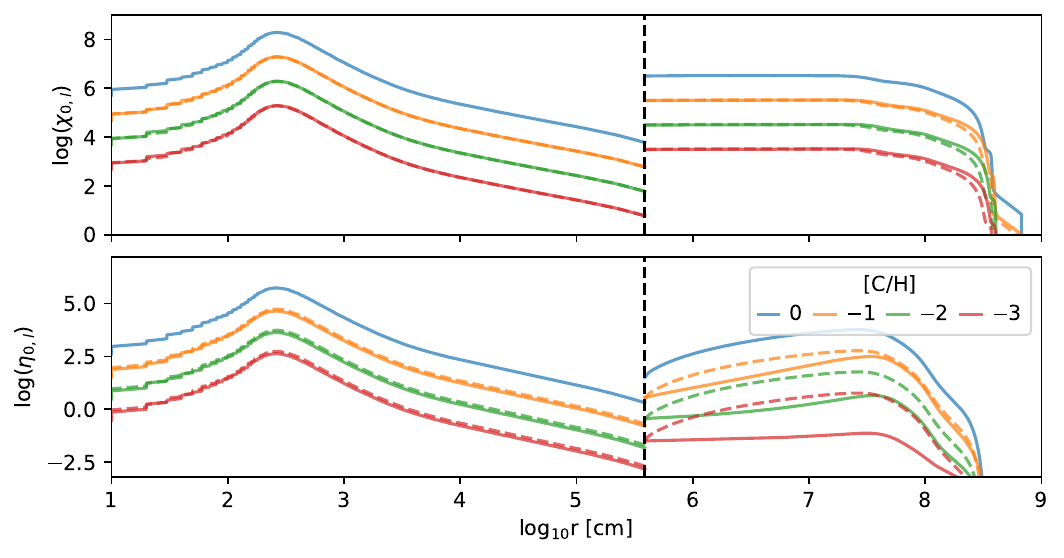}
\caption{The change in radiative properties of the flow due to different carbon abundance. The solid lines are the properties with the abundance as an input in Cloudy. The dashed lines are the values computed by scaling the level population from the results of the [C/H]=0 calculation, demonstrating the non-linear relationship between emissivity and abundance. The units are the same as in Fig.~\ref{fig:line_properties}.
\label{fig:diff_abund}}
\end{figure*}

Observations of chemical species in protoplanetary disks suggest that the elemental abundances in the disks may be modified from the initial values due to processes such as dust trapping, grain growth, or planet formation. Variations in carbon abundance in the (inner) disk may have a non-linear effect on the profiles of the lines forming in the accretion flow/shock. Therefore, we include the abundance as another parameter in our model.  Similar to the previous subsection, we show the line-center optical depth and the line emissivity for an example model of different abundance (compared to the Sun) in Figure~\ref{fig:diff_abund}. The solid lines are for models with the abundance as a parameter in the Cloudy run, whereas the dashed lines show the results when the level populations are scaled from the [C/H]=0 case by the abundance. There is no noticeable difference in the postshock for the two methods, suggesting that the formation of the {\civ} line in the postshock is optically thin, i.e., the curve of growth is in the linear regime. On the other hand, the preshock shows an order of magnitude or more difference in the emissivity when compared between the two methods, suggesting that the line forms in the high optical depth regime.
 
\subsection{C IV Line Profiles} \label{ssec:civ_profiles}
 
\begin{figure*}[t]
\epsscale{1.15}
\plotone{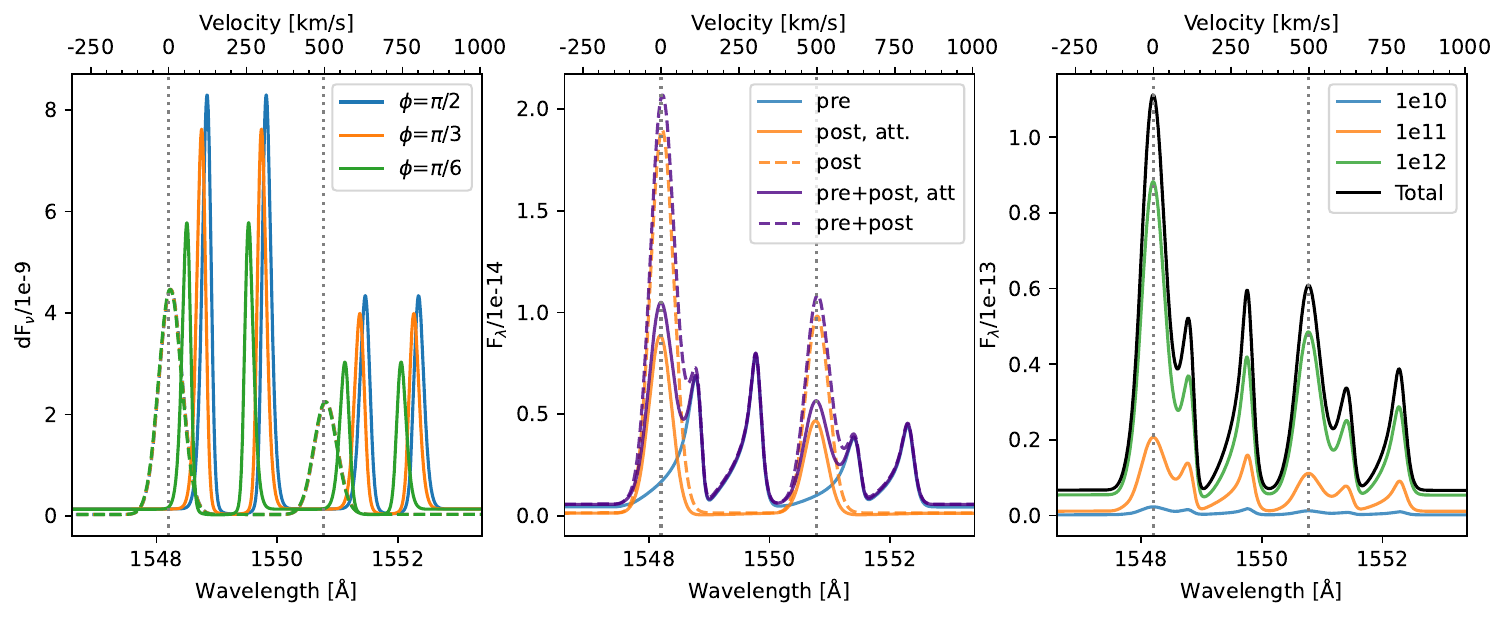}
\caption{The contribution to the line profiles of an example model with $f=0.01$ for all three values of $\curf$, $i=60\degr$, $v_{t, pre/post}=50\,\kms$, [C/H]$=-1$, and distance of 360\,pc. The dotted lines mark the rest-frame line centers.
\emph{Left:} The contribution to the flux from each $\phi$. The solid lines are from the preshock, and the dashed lines are from the postshock. Note that the observer is at $\phi=\pi/2$.
\emph{Center:} The contribution of the postshock and preshock for $\curf=10^{11}\,{\rm erg\,s^{-1}\,cm^{-2}}$.
The dashed lines show the individual emission and their sum, ignoring the attenuation effect of the preshock on the postshock. The spectra plotted with the dotted lines take into account the attenuation of the postshock and the opacity of the preshock.
\emph{Right:} The contribution to the overall line profiles from each energy flux.
\label{fig:components}}
\end{figure*}
 
\subsubsection{Relative Contribution} \label{sssec:contribution}
 Figure~\ref{fig:components} shows the contribution to the line profiles from each component of the flow structure. As seen in the left panel, the low-velocity components originate from the postshock, while the high-velocity, broad components are largely from the preshock. Nevertheless, a significant opacity in the preshock flow is still present at the wavelength of the narrow component, leading to the reduction in flux coming from the postshock. For a given spot ($\theta$, $\phi$), the postshock emission follows essentially a Voigt profile. In contrast, the preshock emission is strongly self-absorbed, and only the wings of the line contribute to the emission (see also Figure~\ref{fig:line_2d}). The smooth, broad, high-velocity component is the superposition of multiple self-absorbed lines at different observed radial velocities from different $\phi$. Therefore, in our model, the final line profile is sensitive to the choice of geometry.

The line center of the narrow component is slightly redshifted, comparable to observations \citep{ardila2013}. However, there is very little difference in line flux and shape from different $\phi$. This is because the terms $\cos\theta_0$ cancel out each other in Equation~\ref{eq:fnu} and Equation~\ref{eq:inu}. The low opacity/optical depth in front of the emitting region (near the heated photosphere; Figure~\ref{fig:line_2d}) makes the exponential term in Equation~\ref{eq:inu} insensitive to the viewing angle $\theta_0$.
 
The center panel of Figure~\ref{fig:components} shows the relative contribution to the overall line profile from the postshock and the preshock for $\curf=10^{11}\,\escm$. The fluxes of the narrow components are attenuated by the preshock, especially at the red side of the line. This attenuation comes from the blue side of preshock's line and the continuum. Depending on the inclination, the attenuation from the preshock may be significant, up to $\sim50\%$ in some cases.
 
The total line profile of the doublet is shown in the right panel of Figure~\ref{fig:components}. As mentioned earlier, the accretion flow is split into five flows (Figure~\ref{fig:geometry}), with the low and moderate energy-flux flows split into the top and bottom components. To illustrate the relative importance of each flow, the filling factor for each energy flux is set to be $f=0.01$. It is evident that the high energy flux $\curf=10^{12}\,\escm$ contributes the most to the line profile for the same $f$. On the other hand, the lowest energy flow contributes the least for the same $f$. For the model of a real star, however, the relative contributions of the components will be different since the relative values of the filling factors are different.
 
As noted earlier, we assume that the outer, lower-density parts surrounding the inner, denser part of the accretion funnel do not add much opacity and will not affect the emerging {\civ} profiles of the denser part. Since the shock locations depend on the flow's ram pressure and the ambient photosphere, different flows are expected to form the shock at different locations. Nevertheless, how these differences affect the {\civ} profiles is beyond the scope of this paper, as this requires full 3D modeling.
 
\subsubsection{Geometry Effects} \label{sssec:inclination}
As mentioned earlier, the overall line shape strongly depends on the geometry. The information is encoded in the values of $\theta_0$, which depend on the relative values of inclination $i$ and the co-latitude of the flows $\theta$ (c.f., Figure~\ref{fig:geometry}). Again, we consider the structure of the flow and the filling factors as in Section~\ref{sssec:contribution}. For $R_i=3.0R_*$, we get a co-latitude of $\sim35\degr$. 
 
\begin{figure*}[t]
\epsscale{0.9}
\plotone{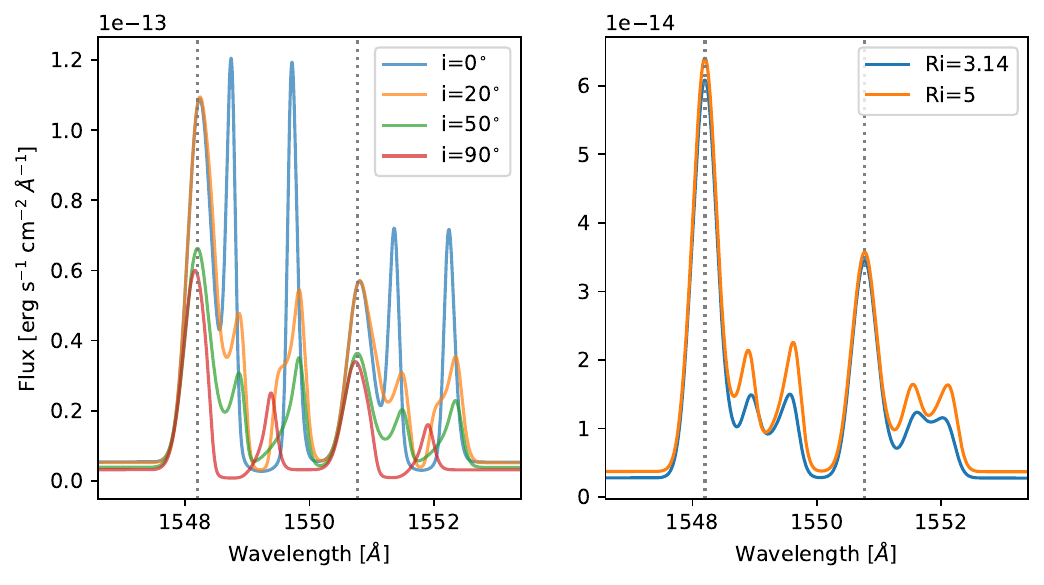}
\caption{\emph{Left}: The total line profile for an example model as seen at different inclinations (left) while other parameters are the same as in the right panel of Fig.~\ref{fig:components}. \emph{Right}: An example of a model profile calculated at different magnetospheric truncation radii and otherwise the same accretion/geometric parameters. The dotted lines mark the rest-frame line centers.
\label{fig:inclination}}
\end{figure*}
 
Figure~\ref{fig:inclination} shows the total line profiles of the same model for different viewing inclinations. Both the narrow and broad components shift to the blue as the inclination increases. For $i<\theta$, the fluxes of the narrow components are high and barely change. On the other hand, the broad components show a significant reduction in flux between $i=0\degr$ and $20\degr$. This is because, at low $i$, a wider range of $\phi$ is seen, and they have a similar $\cos\theta_0$. This leads to a generally higher flux at low $i$ as more surface area is seen. A similar $\cos\theta_0$ leads to the postshock line being in a narrow range of velocities. Therefore, the blue wings of the line in the preshock do not have much overlap over the postshock, so the emission from the postshock (which is insensitive to $\cos\theta_0$; Section~\ref{sssec:contribution}) is largely unaffected. Due to high optical depth, the preshock is affected by a slight change in $\cos\theta_0$, resulting in a significant change in fluxes. 
 
For $i>\theta$, the flux in the preshock component continues to decrease with increasing $i$. The velocity shift of the preshock emission is also more prominent, such that at $i=90\degr$, the blue wing merges with the emission coming from the postshock and boosts the narrow component's flux. The postshock component is unaffected by the optical depth effects, and the flux changes are due mostly to different coverage in $\phi$.
 
The right side of Figure~\ref{fig:inclination} shows the changes in the line profiles for different truncation radii while other parameters are fixed. Only a slight change in the overall line fluxes is present, but the line shape remains very similar. This suggests that the shape of the profiles is more affected by the apparent shape of the emission region (controlled by the inclination) rather than the region's size or location on the star (controlled by the truncation radius).

\section{Applying the Model to ULLYSES Orion Targets} \label{sec:orion}
We test whether the {\civ} line originates in the accretion shock by applying our model to a set of contemporaneous UV to infrared spectroscopic observations. These data were acquired through the \emph{Hubble} Ultraviolet Legacy Library of Young Stars as Essential Standards Director's Discretionary program (ULLYSES; \citealt{roman-duval2020}, along with community-led programs Outflows and Disks around Young Stars: Synergies for the Exploration of Ullyses Spectra (ODYSSEUS; \citealt{espaillat2022b}) and  PENELLOPE \citep{manara2021}. These programs observed a large sample of T Tauri stars across different mass ranges, ages, and mass accretion rates. These observations include medium-resolution FUV spectra (\textit{HST}/COS), low-resolution NUV/Optical spectra (\textit{HST}/STIS), as well as contemporaneous ground-based high/moderate-resolution spectra from the ESO's VLT. The data are public and are accessible through the Mikulski Archive for Space Telescopes (MAST) \footnote{Website: \url{https://mast.stsci.edu/search/ui/\#/ullyses}}.
 
The first set of ULLYSES stars that were studied were those in the Orion OB1 and $\sigma$ Ori regions \citep{manara2021,pittman2022}. We selected a subset of these stars, namely CVSO~104, CVSO~107, CVSO~146, CVSO~176, CVSO~58, CVSO~90, and CVSO~109A, to be included in our study since their accretion properties have been characterized using accretion shock models \citep{pittman2022}. The FUV spectra analyzed in this paper are obtained directly from the ULLYSES Public Data Release 6 (DR6). The COS spectra are binned at 3 pixels but are not corrected for reddening as we also fit for the {\av} (see Section~\ref{sssec:orion-fitting-procedure}).
 
The study provides most of the required stellar and accretion parameters for our modeling, leaving only the truncation radii unknown. Optical spectra, specifically of the {\halpha} and {\hbeta} lines, are needed to determine a truncation radius $R_i$ using the magnetospheric accretion model. We use contemporaneous spectra observed by the PENELLOPE program \citep{manara2021} and apply the magnetospheric flow model of \citet{muzerolle2001} to determine the truncation radius of each star.

\subsection{Determining Disk Truncation Radius} \label{ssec:trunc_rad}
The truncation radius is an important input for the shock model and our line model; it is typically assumed to be $5\,R_*$ \citep[e.g.,][]{gullbring1998,calvet1998}. However, recent studies have shown evidence of this radius being less than $5\,R_*$ (e.g., \citealt{espaillat2022b,thanathibodee2023}). Since our model relies upon the different infall velocities among different accretion flows, better estimates of the truncation radii are needed. We note, however, that the free-fall velocity only varies with the truncation radius as $v_s\propto\sqrt{1-R_*/R_i}$, so we can afford a large uncertainty in $R_i$. 
 
Here, we use the magnetospheric accretion flow model of \citet{hartmann1994,hartmann1998} and \citet{muzerolle2001} to fit the hydrogen Balmer H$\alpha$, H$\beta$, and H$\gamma$ lines of each star. The model's parameters include the mass accretion rate ($\Mdot$), truncation radius ($R_i$), accretion flow width at the disk ($W_r$), maximum flow temperature ($T_{max}$), and inclination ($i$). For each star, we create a grid of models varying the mass accretion rates by a factor of a few around those found by the shock model to account for the unknown calibration of accretion rate measurements between the two models. Specifically, we vary $\Mdot$ between $1-5\times10^{-8}\,\msunyr$ for CVSO~58, CVSO~90, CVSO~104, and CVSO~107, between $1-8\times10^{-8}\,\msunyr$ for CVSO~176, and between $0.2-1.1\times10^{-8}\,\msunyr$ for CVSO~146. We adopt wider ranges for other parameters. Depending on the star, the grid size ranges from 6000 to 18000 models. Following the procedure outlined in \citet{espaillat2022b}, we fit the resulting model profiles with the observation, taking into account the instrumental resolutions, by calculating the $\chi^2$. The final model parameters of a given star are calculated from the weighted average parameters from each epoch/instrument and are shown in Table~\ref{tab:flow_model_results}. Appendix \ref{sec:balmer_line_fit} shows the details of the fitting and the results for individual lines. For CVSO~109A, we adopted the flow best-fit parameters from \citet{espaillat2022b}.

\subsection{Redetermining the Accretion Flow's Filling Factors} \label{ssec:orion-filling_factor}
The filling factors of the targets determined by \citet{pittman2022} assumed a disk truncation radius of 5\,R$_*$ for all but CVSO~109A. For self-consistency, we followed the same procedure outlined in that work to redetermine the filling factors of the accretion flows using the new truncation radii. In short, we calculated the \cite{calvet1998} accretion shock models for each star given its stellar parameters and $R_i$ determined according to Section~\ref{ssec:trunc_rad}. We then use an MCMC process to fit the models to the \textit{HST}/STIS NUV--optical continua by varying the filling factors $f_{h,m,l}$ and V-band extinction {\av}. The re-determined mass accretion rates, filling factors of the flows, and {\av} are listed in Table~\ref{tab:flow_model_results}. The results are slightly different from that of \citet{pittman2022}. Most notably, the accretion rates are higher in this work than in the previous work. This is to be expected since the luminosity of the accretion column $L$ scales as $L\propto(1-R_*/R_i)L_{acc}\propto(1-R_*/R_i)\dot{M}$ \citep{calvet1998}. Therefore, a smaller {\ri} than the previously assumed 5\,$R_*$ would lead to larger {\mdot} for the same observed $L$.
 
\begin{deluxetable*}{lCCCCCCCCCC}
\tablecaption{Results of the Accretion Flow and Shock Models \label{tab:flow_model_results}}
\tablehead{
\colhead{Star} &
\colhead{\mdot$_{\text flow}$} &
\colhead{\ri} &
\colhead{\rw} &
\colhead{\tmax} &
\colhead{$i$} &
\colhead{\mdot$_{\text shock}$} &
\colhead{$f_{\text 1E10}$} &
\colhead{$f_{\text 1E11}$} &
\colhead{$f_{\text 1E12}$} &
\colhead{$\mathrm{A_V}$\tablenotemark{a}}
\\
\colhead{} &
\colhead{($10^{-8}\,\msunyr$)} &
\colhead{(R$_{\star}$)} &
\colhead{(R$_{\star}$)} &
\colhead{($10^3\,$K)} &
\colhead{(deg)} &
\colhead{($10^{-8}\,\msunyr$)} &
\multicolumn{3}{c}{(fraction of stellar surface)} &
\colhead{(mag)}
}
\startdata
CVSO~58 & 2.6\pm0.1 & 2.6\pm0.5 & 0.2\pm0.0 & 7.8\pm0.4 & 61\pm3  & \pmm{2.10}{0.12}{0.11} & 0.069 & 0.00033 &   0.0116 & \pmm{1.70}{0.04}{0.04}   \\
CVSO~90 & 2.5\pm0.3 & 4.8\pm0.1 & 0.2\pm0.0 & 8.4\pm0.1 & 41\pm1  & \pmm{2.00}{0.04}{0.05}  & \nodata & 0.00026  &  0.0290 & \pmm{1.05}{0.01}{0.01} \\
CVSO~104 & 1.6\pm0.6 & 2.8\pm0.6 & 0.3\pm0.2 & 9.1\pm0.8 & 57\pm11  & \pmm{1.13}{0.01}{0.01} & 0.0479 & 0.00454 & 0.000316 & \pmm{0.01}{0.00}{0.00}\\
CVSO~107 & 2.2\pm0.1 & 2.7\pm0.4 & 0.3\pm0.3 & 7.2\pm0.2 & 54\pm4  & \pmm{2.59}{0.11}{0.11} & 0.0008 &  0.0156  &  0.0049  & \pmm{1.02}{0.03}{0.03} \\
CVSO~109A\tablenotemark{b} & 2.6\pm1.3 & 2.2\pm0.2 & 0.2\pm0.0 & 7.2\pm0.0 & 37\pm8 & \pmm{4.01}{0.13}{0.23} & 0.0008 &  0.0217  & 0.00167 & \pmm{0.21}{0.02}{0.05} \\
CVSO~146 & 0.5\pm0.0 & 2.9\pm0.3 & 0.3\pm0.1 & 9.3\pm0.2 & 65\pm5  & \pmm{2.20}{0.08}{0.07} & 0.149 &  0.0094  & 0.00142 & \pmm{0.91}{0.03}{0.03}\\
CVSO~176 & 2.9\pm0.1 & 2.5\pm0.5 & 0.2\pm0.0 & 8.2\pm1.1 & 54\pm7  & \pmm{0.911}{0.016}{0.016} &  0.0373 &  0.0006 & 0.000046 & \pmm{0.49}{0.02}{0.02}  \\
\enddata
\tablenotetext{a}{The extinction used in the shock model assumes the \citet{whittet2004} law.}
\tablenotetext{b}{The flow parameters are from \citet{espaillat2022b}.}
\end{deluxetable*}
 
\subsection{Fitting the C IV line} \label{ssec:orion-fitting}
\subsubsection{Grids of Models} \label{sssec:orion-fitting-grid}
For a given star, we create a grid of models varying the pre/postshock turbulent velocities from 5 to 100$\kms$, the carbon abundance [C/H] from -3.0 to 0.0, and the inclination ranging between those measured with the $v\sin{i}$ \citep{pittman2022} and the flow modeling (Section~\ref{ssec:trunc_rad}) with a step of 5 degrees. The geometry, which is controlled by the truncation radius ({\ri}) and the filling factors ($f_{l,m,h}$), is fixed. We assume that the abundance is constant throughout the structure and that the preshock and postshock turbulence are independent. The final model profiles are convolved to the line spread function (LSF) of COS and re-binned to the same wavelength scale as the observations. Depending on the range of possible inclinations in each star, there are 10,400 to 41,600 models per grid.

\subsubsection{Fitting procedure}\label{sssec:orion-fitting-procedure}
The COS observations need to be corrected for the intrinsic radial velocity. Since there are no photospheric lines in the FUV spectra, we use {\htwo} lines to determine the radial velocity since they form in the disk \citep{herczeg2004, france2012, hoadley2015, gangi2023}, and the line center should be at zero velocity compared to the star. For each line, we fit a Gaussian and determine the velocity shift of the line compared to the line center at rest. We use eight strong lines for these fits, at 1446.120\,{\AA}, 1467.08\,{\AA}, 1489.570\,{\AA}, 1500.450\,{\AA}, 1504.760\,{\AA}, 1524.650\,{\AA}, 1556.840\,{\AA}, and 1562.410\,{\AA}, and the final radial velocity is the mean of the fits. We then correct the spectra with these RVs. The fitted radial velocities for the stars are between $\sim10-20\,\kms$, which is comparable to the wavelength uncertainty of the COS spectrograph at $\sim10\,\kms$.

The blue side of the {\civ} doublet overlaps with a fluorescent {\htwo} emission line at 1547.34 \AA, which we account for in our fitting. We assume that the feature can be represented by a Gaussian line profile \citep{france2012}; however, line blending makes it challenging to directly measure its best-fit parameters. Instead, we first determined the best-fit Gaussian amplitude, FWHM, and central velocity of the well-separated {\htwo} emission line at 1489.57 \AA, using an interactive model fitting routine that convolves the ``intrinsic" Gaussian with the wavelength-dependent line-spread-function of the \textit{HST}-COS instrument \citep{arulanantham2018}. This feature originates from the same upper level and corresponding disk gas population as the 1547.34 \AA \, transition $\left([v', J'] = \left[1, 4 \right] \right)$. Since UV-fluorescent {\htwo} emission lines are primarily Keplerian-broadened, with flux ratios that scale with the ratio of Einstein $B$ coefficients, we assume that the two features share a Gaussian FWHM and central velocity. We then multiply the best-fit amplitude of the 1489.57 \AA \, line by the ratio of Einstein coefficients of the two transitions to estimate the Gaussian amplitude of the 1547.34 \AA \, emission line.
 
Since the continuum calculated from the model only includes the contribution from the shock emission (the star's photosphere is negligible), it does not represent the real observation, which may include other sources, such as the unknown chromosphere and the underlying molecular dissociation continuum \citep{france2017,france2023,espaillat2022b}. Therefore, we only fit the residual profile by subtracting the continuum in the observation and the model. We also consider the possibility that the chromosphere contributes to the {\civ} line emission. \citet{ardila2013} presents the line of a number of non-accretion young stars, showing that the profiles are narrow and weak. We calculate the flux per stellar surface area around $(2\pm1)\times10^5\;\mathrm{erg}\;\mathrm{s}^{-1}\;\mathrm{cm}^{-2}$ by averaging across four stars (RECX~1, TWA~7, LKCa~17, and HBC427) using stellar parameters from \citet{ingleby2013}. Multiplying by the surface area of the stars we model in this study, we expect that the chromosphere contributes only $0.7-7\%$ to the total {\civ} flux we observed, thus considering it negligible.

Given that the extinction at 1550{\AA} is high, a slight change in {\av} may significantly change the flux level. Therefore, we also allow {\av} to vary between the range of those calculated from the shock modeling (Section~\ref{ssec:orion-filling_factor}, Table~\ref{tab:flow_model_results}) and the values from fitting the optical spectra \citet{manara2021}. We expand the {\av} range by 0.1 magnitude on each side to account for any possible uncertainty. For model $i$, the fitting flux can be written as
\begin{equation}
    F_{\lambda, i, fit} =  10^{-\frac{A_VA(\lambda)/A(V)}{2.5}}F_{\lambda, i, 0} + F_{\lambda, H_2},
\end{equation}
where  $F_{\lambda, H_2}$ is the Gaussian {\htwo} profile determined above. Here, we use $A(\lambda)/A(V)$ from \citet{cardelli1989} with $R_V = 3.1$, which is a fine approximation for the Orion OB1b sub-association the targets belong to \citep{briceno2019}. The best fit {\av} for each model is determined using Scipy's \texttt{curve\_fit} function. We then calculate the $\chi^2$ for each model by
\begin{equation}
\chi^2_{i} = \sum\frac{(F_{j, i, best fit} - F_{j, obs})^2}{\sigma^2_{j, obs}},
\end{equation}
where $j$ indexes over the pixels within $750\,\kms$ of the mean line center at 1549.488\,{\AA}. The best-fit parameters for each star are the averages from the top 100 best fits, representing the top $\sim0.24\%$ to $\sim1\%$ of the model grid, depending on the star.

\subsection{Fitting Results} \label{ssec:orion-results}
 
\begin{figure*}[t]
\epsscale{1.15}
\plotone{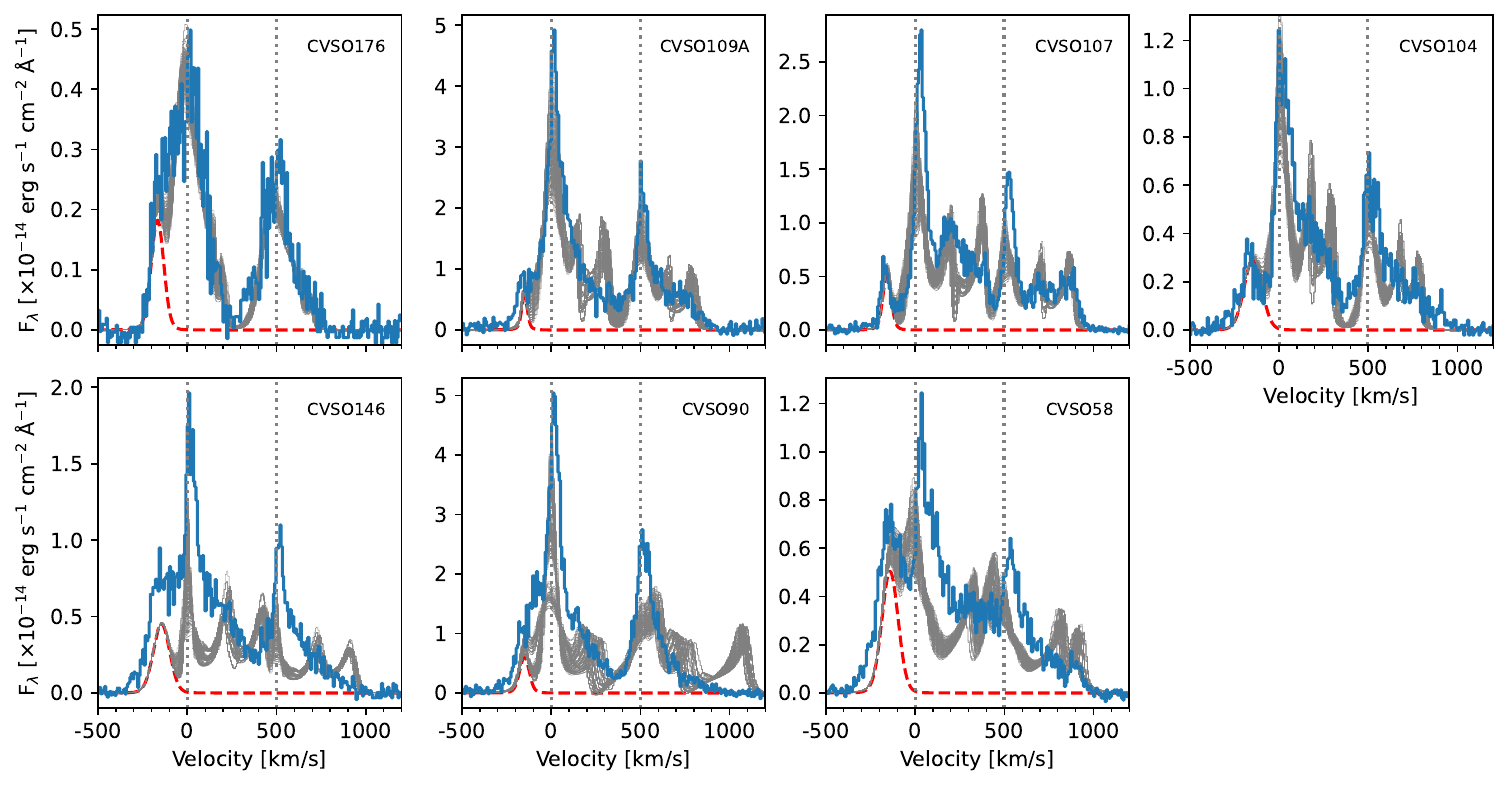}
\caption{The top 100 best fits. The blue lines are the observations. The red dashed lines are the Gaussian representation of the {\htwo} feature. The black lines are the reddened {\civ} models plus the Gaussian. The continua have been subtracted for the models and the observations. The models fit the observations well for stars in the top rows (CVSO~176, CVSO~109A, CVSO~107, and CVSO~104).
\label{fig:best_fits}}
\end{figure*}
 
\begin{figure*}[t!]
\epsscale{1.15}
\plotone{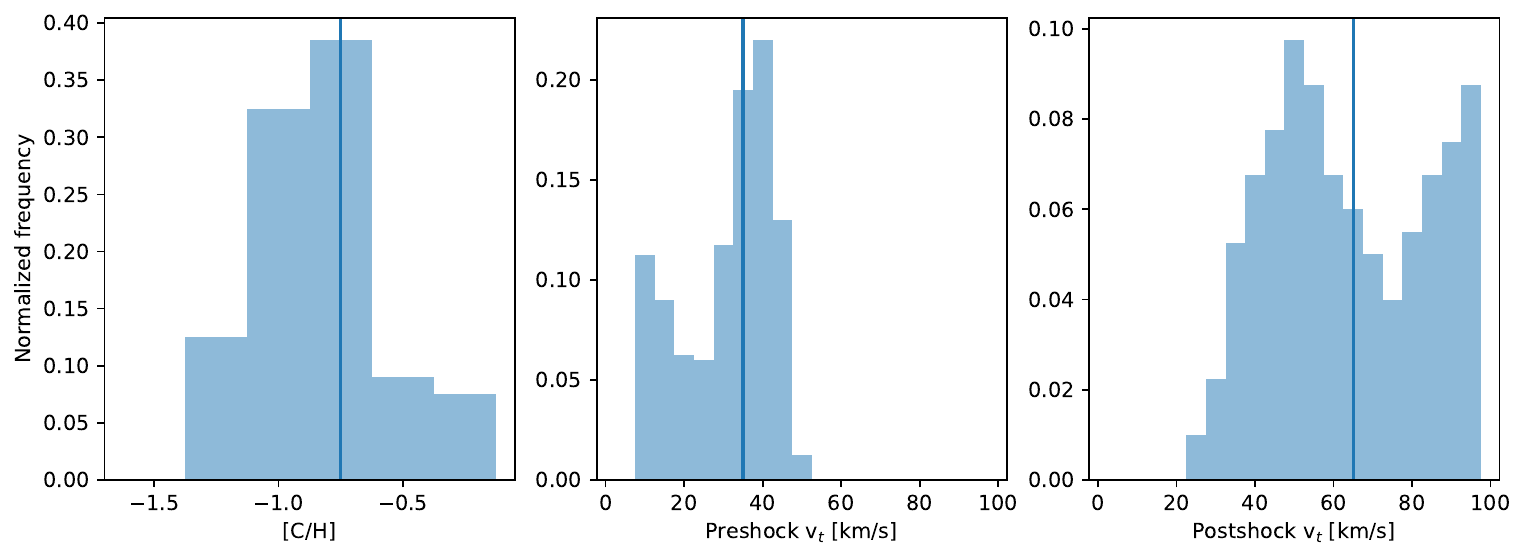}
\caption{The distributions of fitting parameters from the top 100 best fits of CVSO~176, CVSO~109A, CVSO~107, and CVSO~104. The vertical lines mark the median of the distribution.
\label{fig:best_fits_distribution}}
\end{figure*}
 
Figure~\ref{fig:best_fits} shows the top 100 best fits for the sample. The model profiles roughly agree with the observations for CVSO~176, CVSO~109A, CVSO~107, and CVSO~104. In particular, the model qualitatively reproduces the narrow and broad components of the doublet's profiles and the absolute flux levels. We note that the model also reproduces the blue sides of the narrow components and the sharp drop on the red sides of the broad components.

Selecting 100 best fits for CVSO~176, CVSO~109A, CVSO~107, and CVSO~104, we show in Figure~\ref{fig:best_fits_distribution} the combined distributions of the fitted parameters The vertical lines show the median of the distribution: [C/H]=-0.8, $v_{t, pre}=35\,\kms$, and $v_{t, post}=65\,\kms$. These indicate that the accretion flows of these four stars are carbon-poor and highly turbulent. For the preshock, the turbulent velocity is $\sim20\%$ or less of the free-fall velocities, with the peak around $50\,\kms$. On the other hand, the postshock seems to have a bimodal distribution in the turbulent velocity. The first peak at $\sim30\,\kms$, is comparable to the velocity near the base of the postshock (about $90\%$ from the shock front). The second peak near $100\,\kms$ could be the artifact of the fitting where the model tries to compensate for the large width of the narrow component that it cannot reproduce by increasing the turbulence to broaden the line. This is discussed further in Section \ref{ssec:discussion-turb}.
 
\begin{figure*}[t]
\epsscale{0.9}
\plotone{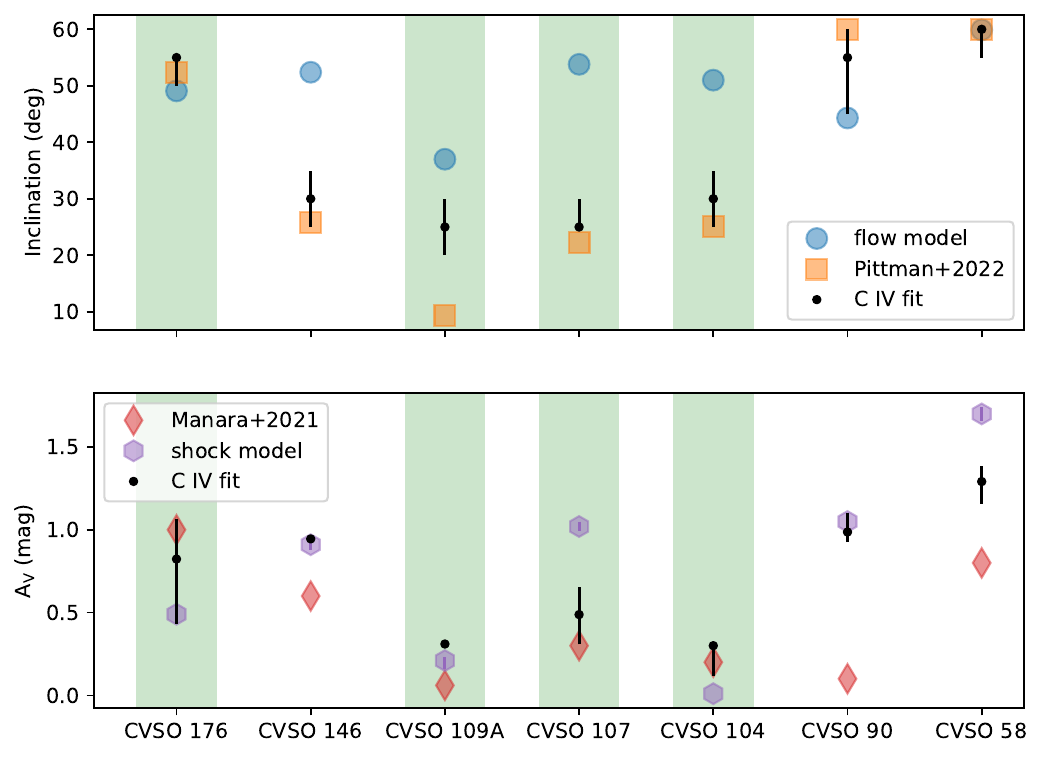}
\caption{The best fits inclination and {\av} compared to the values from the literature and accretion modeling. The error bars are the standard deviations from the top 100 best fits. The green shades mark stars that have good fits. The model inclinations are constrained to be between those from {\vsini} \citep{pittman2022} and from accretion flow modeling (Section~\ref{ssec:trunc_rad}). The ranges of {\av} are from \citet{manara2021} and the shock modeling (Section~\ref{ssec:orion-filling_factor}), plus 0.1 magnitude for uncertainty.
\label{fig:incl_av_comparison}}
\end{figure*}

In Figure~\ref{fig:incl_av_comparison}, we compare the best-fit inclinations and {\av} with their possible ranges. There is a slight preference of inclinations toward the values determined from {\vsini} \citep{pittman2022}. On the other hand, there is no clear preference of {\av} toward either the shock modeling or the optical spectra fitting \citep{manara2021}.

\section{Discussion} \label{sec:discussion}
\subsection{Geometry and Filling factors} \label{ssec:discussion-geometry}
Analyses of a large number of CTTS by \citet{ardila2013} showed that the shapes of the {\civ} doublet were diverse and did not correlate with accretion rates, line luminosity, or inclination. They also did not see any dependence of accretion rates or accretion luminosity on inclination, but the sample size was too small to be definitive. Nevertheless, their data suggested that the emitting region of the {\civ} doublet is inconsistent with either being localized or covering the entire star. That is, it cannot be a small spot or a very large spot. This, however, does not rule out multiple accretion spots distributed across the surface.
 
Our suggested geometry of the accretion ring is consistent with this picture. Specifically, the emission ring should be mostly seen by an observer regardless of inclination. In reality, the ring is likely not axisymmetric, but as long as the {\civ} emission comes from a sufficiently large range of azimuthal angles, {\civ} is always seen in accreting stars. 
 
One major difference between our results and \citet{ardila2013}'s conclusion is that we found that the broad component of the {\civ} doublet is optically thick, similar to what was suggested by \citet{lamzin1998,lamzin2003a}, whereas the narrow component is optically thin. We note that, as pointed out by \citet{lamzin1998}, the flux ratio of 2:1 between the doublet component does not necessarily mean that the emission is optically thin, but it rather reflects the electron's collisional excitation coefficient. Moreover, the self-absorption of the emission coming from an area in our model suggests that most of the local emission in the broad component is not observed, so the total line fluxes should not necessarily correlate with other properties. This is also found to be the case by \citet{ardila2013}.
 
In our model, the broad component's shape depends greatly on the flow's geometric parameters, such as the inclination and truncation radius, as seen in Figure~\ref{fig:inclination}. This is because the broad component is composed of many different wings of high-optical depth lines, the shape of which also depends on the geometry. Our adapted geometry for the accretion hotspots, i.e., axisymmetric rings, may not be able to fully explain the observations of every star in our sample. Rather, the line profiles may originate from a more complex  distribution of small(er) accretion hotspots on the stellar surface as described by \citet{lamzin2003a}. Hence, we plan to explore other geometries, such as banana-shaped hotspots \citep{kurosawa2013}, in future works.
 
There are also indications that, in reality, the flow of the accreted material is not just oriented towards the star as prescribed in a 1D infall model. Instead, the post-shock material hits upon the much higher-density photosphere, and the thermodynamic pressure overcomes the magnetic confinement from the dipolar field. Ionized material then escapes sideways out of the accretion funnel. \citet{orlando2010} see this scenario in magneto-hydrodynamic simulations and \citet{brickhouse2010} invoke it to explain the O~{\sc vii} emission measure observed in X-rays from TW~Hya, which is much larger than expected in the post-shock zone. If this scenario is true for the targets studied here, there can be additional localized, blue-shifted {\civ} emission or absorption (depending on geometry and temperature of the gas spilling out sideways and upwards), which is not represented in our 1D flux model. In particular, in CVSO~146, CSVO~90, and CVSO~58, there is considerable blue-shifted emission that is not seen in our model (Figure~\ref{fig:best_fits}) but might be explained this way.
 
\subsection{Turbulence} \label{ssec:discussion-turb}
While the implied high turbulence velocity may be surprising given that hotspots are observed to have $\sim$kG magnetic field strength \citep[e.g.,][]{johns-krull2013}, it is not clear if that is true for all spots at all times. In fact, observations and models have shown patterns of hotspots on the stellar surface that are more complex than what would be expected from a dipolar infall \citep{gregory2008,donati2011b,donati2011c,donati2024,orlando2010,sacco2008}. Three-dimensional simulations, while not complete, have discovered many different modes of accretion with complex accretion structures \citep[e.g.,][]{romanova2012,romanova2021,zhu2024}. Therefore, it is possible that what we are seeing as turbulent flows may be the consequence of such complex systems.
 
There is no theoretical prediction for the exact turbulence value to expect.  Observationally, our model fitting strongly prefers $50\,\kms$, and this figure is not inconceivable for the preshock with a flow velocity of a few hundred $\kms$. Furthermore, slight variations of the radius where the material is loaded onto the field lines already lead to velocity variations on the 10\,$\kms$ scale. While the postshock's flow speed and sound speeds are lower than the supposed turbulent velocity, the plasma in the region is dominated by hydrodynamic pressure, and the flow can be deflected to the side and is observed as turbulent flow as shown in 2D/3D simulations \citep{orlando2010,bonito2014}. In this case, the turbulent velocity can be larger than the radial flow speed or the sound speed.

Alternatively, it is also possible that the fitted turbulent velocity is artificially high due to the broad observed velocity profile of the narrow component compared to the calculated model at a given position on the star. In our model, the narrow component forms in the transition region between the postshock and the heated photosphere, which has very low to zero velocity. If, instead, the line forms in the upper part of the postshock, with velocity up to $v_{\rm freefall}/4$, then the turbulent velocity does not need to be high for the model to fit the observation since the narrow component of the model profile can be broadened by the wide distribution of the observed angles. As the line forms at temperatures of $\times~10^5$\,K (c.f., Figure~\ref{fig:flow_properties}), this could indicate the postshock cools faster than (or is not as hot as) our shock model. This can be made possible if the shock lands at an angle (instead of perpendicular) to the surface, and the energy can dissipate to the side.
 
Another possibility is that the fitted turbulence is, in fact, reflecting a range in the observed radial velocity coming, for instance, from horizontal velocity components of the flows observed at an angle. Therefore, this extra velocity component will help shift the emerging line in velocity space for accretion flows on the side of the stellar disk, making the observed line broader. This extra velocity component is not possible in our perpendicular flow model, but it is possible in slanted accretion flows.
 
\citet{burdonov2020} conducted laboratory experiments to study how the accretion flow/accretion shock would behave in the case of a slanted flow compared to a perpendicular flow. They found that slanted flows result in extra material being ejected to the side, creating large perturbations and decreased heating. This scenario is consistent with our results. Specifically, the high turbulent velocity in our model results could indicate strong perturbation in the flow or from spreading material. Decreased heating would also cause the {\civ} line to form higher in the flow where the velocity is still high, making the line spread more prominent.
 
\subsection{Carbon Abundance}\label{ssec:discussion-C_abundance}
Elemental photospheric abundances of young stars derived from their optical/infrared spectra are consistent with solar or close-to-solar values \citep{biazzo2011,dorazi2011,spina2017}. Photospheric abundances are available for 5/7 stars in our sample from APOGEE DR16. \citet{jonsson2020} found that CVSO~104, CVSO~107, and CVSO~109 photospheric carbon abundances are consistent with slightly sub-solar values of [C/H]$\sim-0.2$ to $-0.4$. On the other hand, CVSO~90 and CVSO~58 are more metal-poor, with [C/H] of $-1.6$ and $-0.6$, respectively. Since the model line flux relies on the postshock density, temperature structure, and shock geometry, our model simplification and assumptions may prevent us from accurately measuring carbon abundance in accretion flows. Nevertheless, the low carbon abundance required in our models to explain the \civ~profiles of CVSO~104, 107, 109, and 176, as shown in Figure~\ref{fig:best_fits_distribution}, seems to indicate that the flows' carbon abundance is even lower and, thus,  sub-stellar. These results suggest carbon depletion in the inner disk of these stars, compared to that of the whole disk or the initial cloud.
 
Studies of a neutral carbon line in the infrared \citep{mcclure2019,mcclure2020} suggest low carbon abundance in the inner disk of some young stars. These authors have found that in the $\sim2$~Myr Taurus star-forming region, carbon in the inner disk can be depleted by up to a factor of 42 and by a factor of 100 in the 10\,Myr TW Hya system. Our resulting carbon abundances (Figure~\ref{fig:best_fits_distribution}) are consistent with these values, suggesting that the processes that remove or trap carbon from reaching the inner disks have already occurred in the 5\,Myr Orion systems studied here.

\subsection{Resonance Scattering of the C~IV Doublet}
 
Due to its resonant nature, photons near the C~IV doublet can interact with C~IV atoms. During this interaction, an electron in the ground state is excited to the $P$ state, leading to the absorption features of C~IV, referred to as self-absorption in this study. Furthermore, when the excited electron returns to the ground state, it re-emits C~IV emission photons in a different direction, a process known as resonance scattering. Therefore, considering resonance scattering instead of self-absorption enhances the strength of the C~IV doublet in the other line of sight.
 
Theoretically, the flux ratio of C~IV $\lambda$1548 and $\lambda$1550 is 2:1 due to the statistical weights of the $P_{3/2}$ and $P_{1/2}$ states. However, previous studies have reported various flux ratios of C~IV doublet, ranging from 0.6 to 2.0, in objects related to white dwarfs (e.g., planetary nebulae \citealt{pne1,pne2}, symbiotic stars \citealt{sys1,sys2}) since the flux ratio can vary in an optically thick medium for the C~IV doublet. Specifically, \citealt{sys2} determined that a flux ratio of the C~IV doublet less than unity originates from a strong outflow ($> 450\, \rm km\, s^{-1}$), which is comparable to the separation of C~IV $\lambda$1548 and $\lambda$1550 in velocity space. This variation in the flux ratio of metal doublets due to strong outflows was explored by \citep{chang2024} through the Mg~II doublet near 2800 \AA. Consequently, if the vicinity of the shock region has an outflow velocity faster than $400 \, \rm km\, s^{-1}$ or is optically thick for the C~IV doublet, resonance scattering induces additional variations in its flux ratio and line profile. The effect of resonance scattering in the accretion shock of T Tauri stars will be explored in future works.

\section{Summary and Conclusions} \label{sec:summary}
In this paper, we develop a framework for modeling the {\civ} doublet based on the accretion shock model of \citet{calvet1998} and the accretion flow model of \citet{muzerolle2001}. We then applied the model to a selected sample of Orion OB1 ULLYSES targets. We summarize our findings below.
 
\begin{enumerate}
\item In our model, each of the {\civ} lines is composed of narrow and broad components, with emission originating from the postshock and the preshock, respectively. The line is optically thin in the postshock and optically thick in the preshock.
\item The line shape and line flux of the narrow component depends only on the filling factor of the shock as viewed by the observer. On the other hand, the emission from the preshock is self-absorbed, and the shape and flux are sensitive to the geometry, including the shape of the shock region and the viewing inclination.
\item For the same filling factor, the line emission is dominated by the flow with the highest energy flux.
\item Our axisymmetric model can reproduce the line shapes and fluxes of 4 out of 7 stars from the ULLYSES Orion OB1 sample (CVSO~176, CVSO~109A, CVSO~107, and CVSO~104). This suggests that our model's assumption, especially the shock's geometry, could be crucial in the emerging line flux and shape. Other geometries will be explored in future works.
\item Chromospheric {\civ} emission contributes $0.7$-$7\%$ to the total flux in our sample.
\item Both the postshock and preshock flows are turbulent. For the postshock, the high turbulent velocity could be due to extra velocity components not accounted for by the model or the postshock line emission forms at a shallower depth.
\item For the four stars with good fits, we find that the carbon abundances in the flows are sub-solar and cannot be explained by the photospheric [C/H] of the stars. The results imply that the inner region of disks surrounding these four 5\,Myr stars are carbon-depleted. This is consistent with
    inner disk carbon abundances found with    different measurement methods. 
\end{enumerate}

\begin{acknowledgments}
\noindent
The authors thank the referees for their suggestions, which improved the manuscript. We thank the PENELLOPE team, particularly Carlo Manara, Laura Venuti, and Michael Siwak, for reducing the VLT data used in this work. We also thank Fred Walter for his insightful comments and suggestions.
 
This work is supported by \textit{HST} AR-16129 and benefited from discussions with the ODYSSEUS team (\url{https://sites.bu.edu/odysseus/}); see \cite{espaillat2022b} for an overview of the ODYSSEUS survey.
 
Based on observations obtained with the NASA/ESA Hubble Space Telescope, retrieved from the Mikulski Archive for Space Telescopes (MAST) at the Space Telescope Science Institute (STScI). The specific observations analyzed can be accessed via \dataset[https://doi.org/110.17909/t9-jzeh-xy14]{https://doi.org/10.17909/t9-jzeh-xy14}. STScI is operated by the Association of Universities for Research in Astronomy, Inc. under NASA contract NAS 5-26555.
 
This work is based on observations collected at the European Southern Observatory under ESO programmes 106.20Z8.001, 106.20Z8.002, and 106.20Z8.009; see \citet{manara2021} for an overview of the data set.
 
This research made use of Astropy,\footnote{http://www.astropy.org} a community-developed core Python package for Astronomy \citep{astropy2013,astropy2018}.
\end{acknowledgments}
 
\facilities{ESO:VLT (UVES, X-shooter, ESPRESSO)}

\software{Cloudy \citep{ferland2017}; Astropy \citep{astropy2013,astropy2018}}

\bibliography{tts}{}
\bibliographystyle{aasjournal}
 
\appendix
\restartappendixnumbering
\section{Fitting Balmer Lines with the Magnetospheric Flow Model} \label{sec:balmer_line_fit}
\begin{figure*}[ht]
\epsscale{1.15}
\plottwo{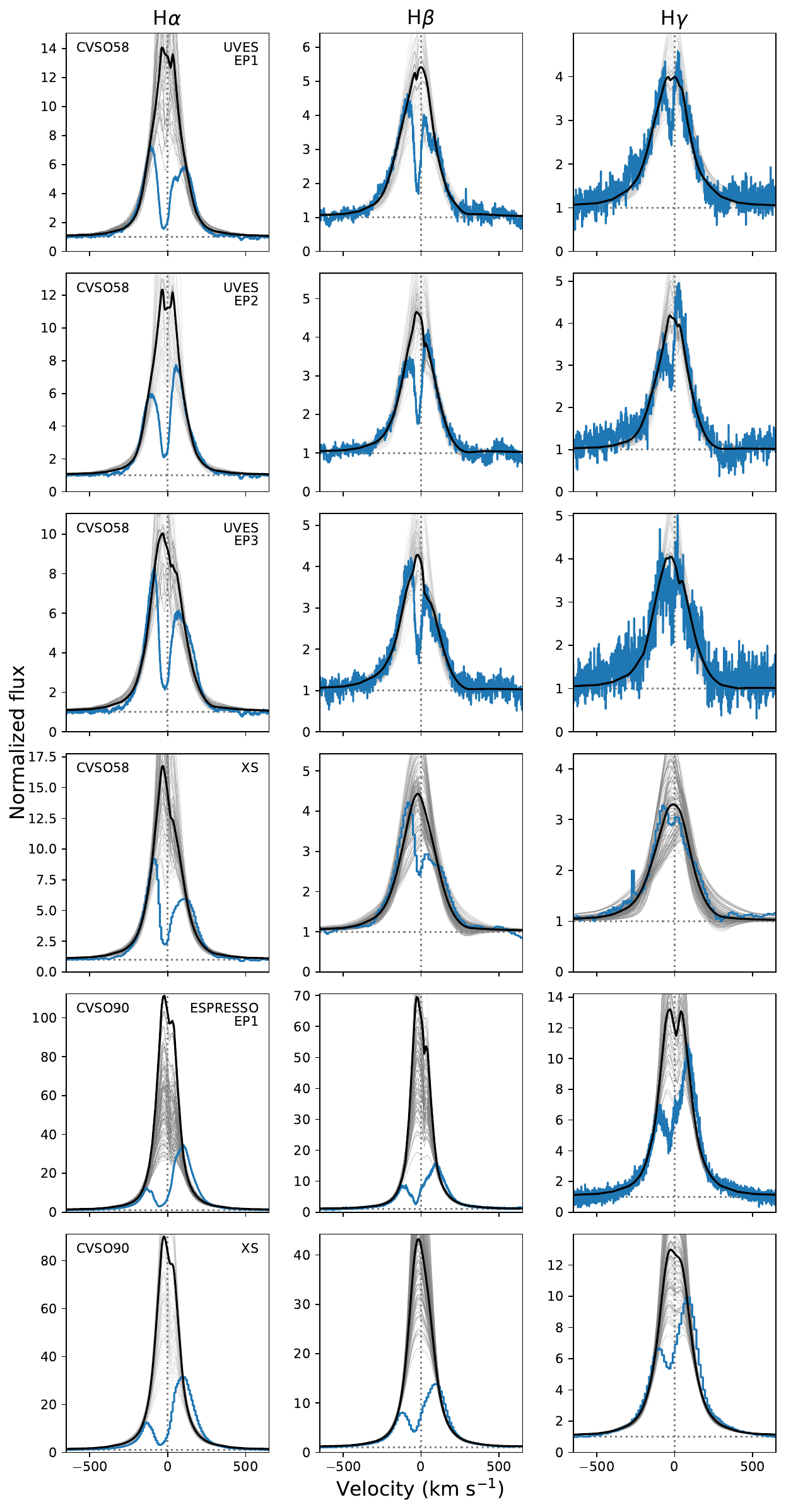}{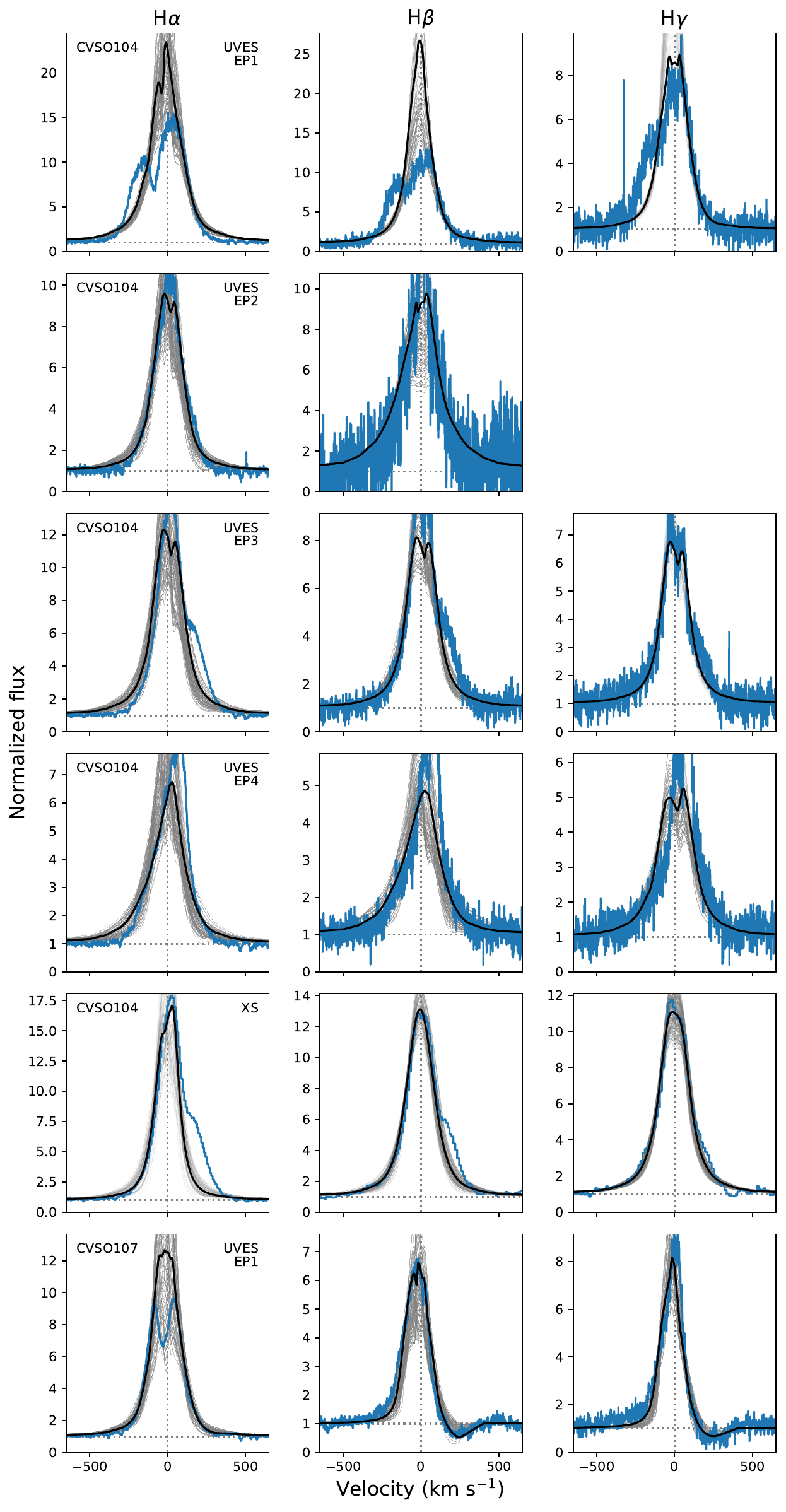}
\caption{The magnetospheric accretion flow model fits to individual Balmer lines. The blue lines are the observations, the grey lines are the 100 best fits, and the black lines are the average of the model profiles. The vertical/horizontal dotted lines mark the line center and the continuum, respectively. The corresponding fit parameters are shown in Table~\ref{tab:flow_model_results_individual}
\label{fig:flow_fit1}}
\end{figure*}

\begin{figure*}[ht]
\centering
\includegraphics[width=0.49\textwidth]{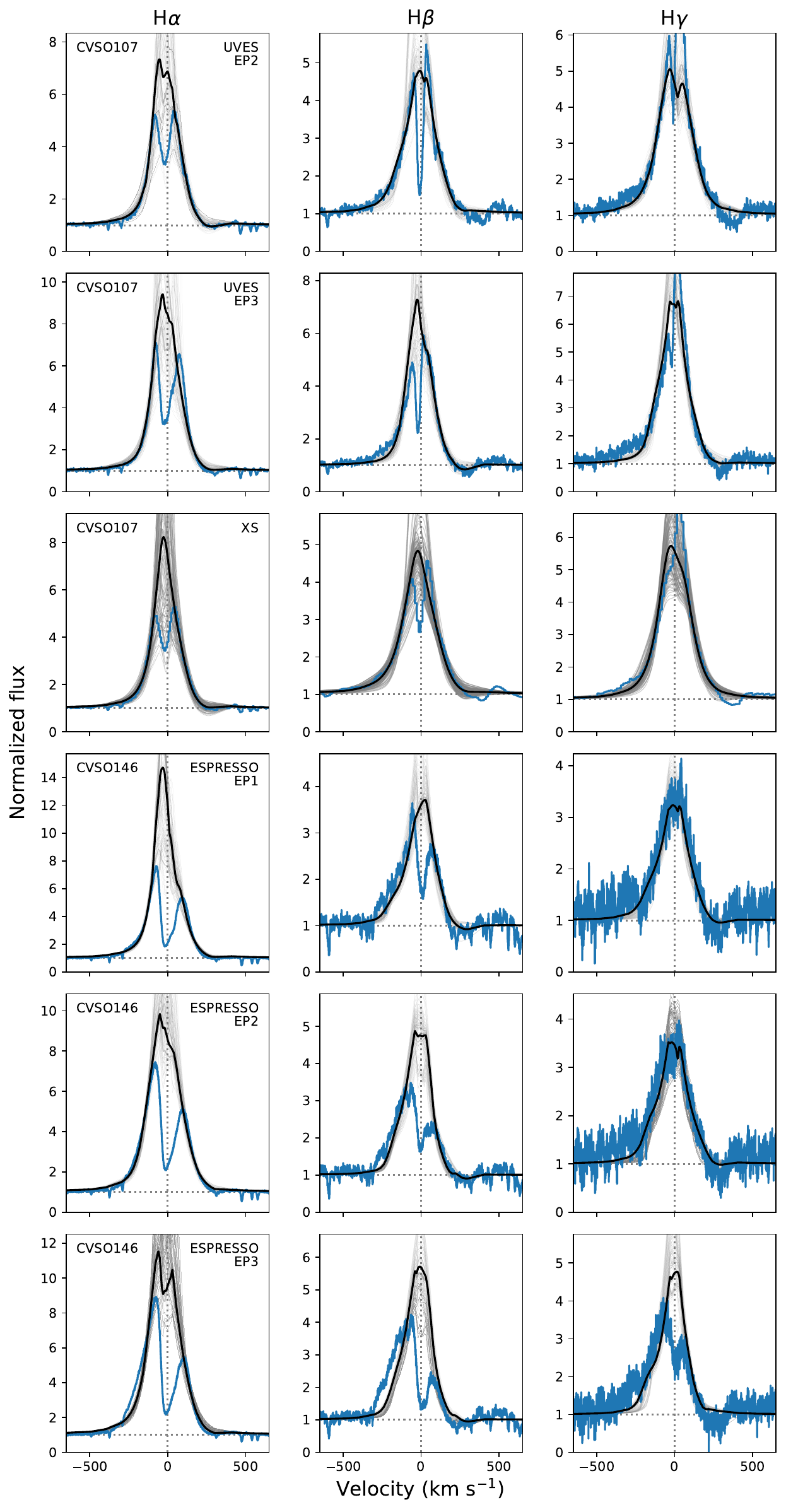}
\includegraphics[width=0.49\textwidth]{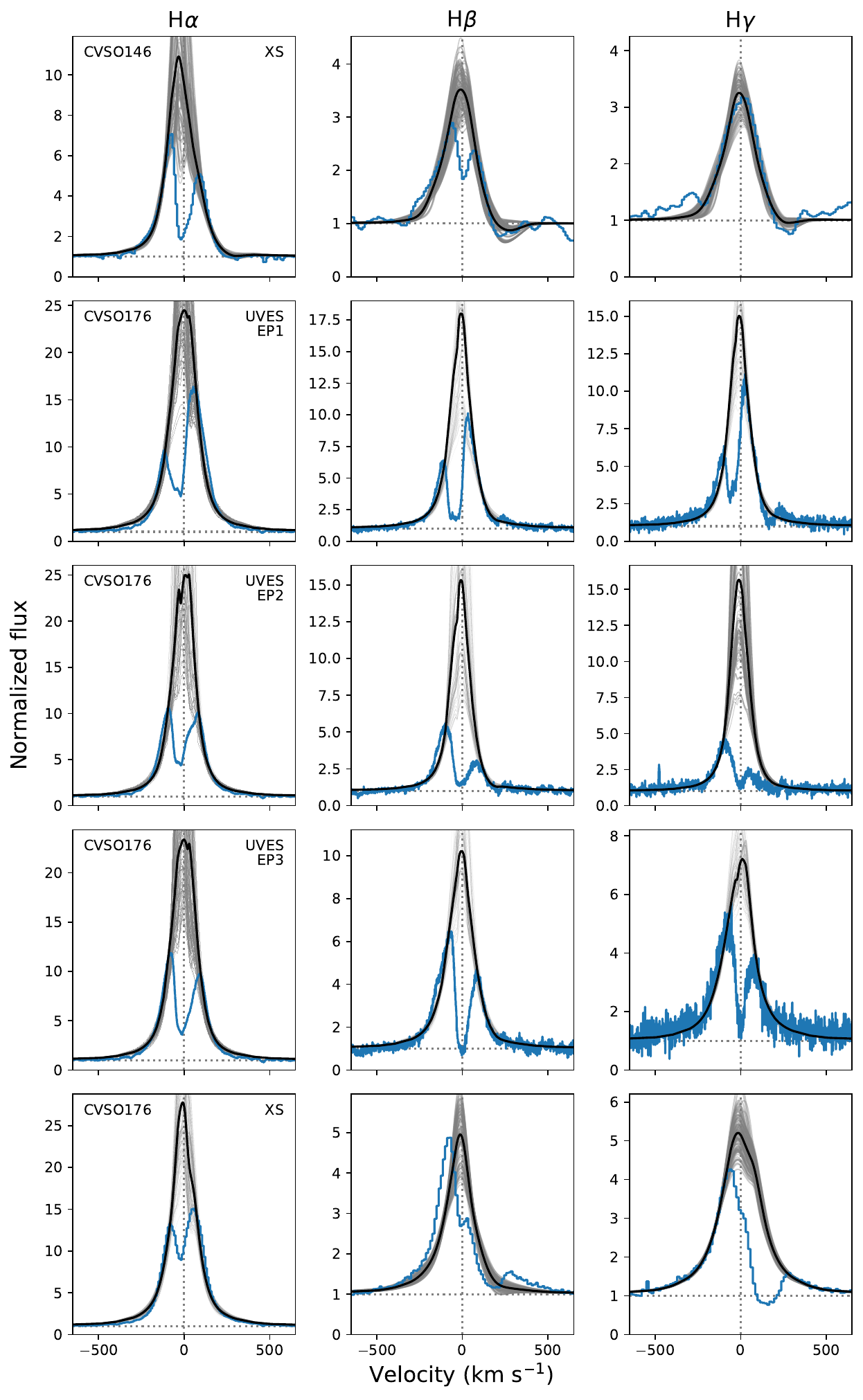}
\caption{Same as Fig.~\ref{fig:flow_fit1}.
\label{fig:flow_fit2}}
\end{figure*}

To estimate the truncation radius of each star, we use the magnetospheric flow model of \citet{muzerolle2001} to fit the {\halpha}, {\hbeta}, and {H$\gamma$} lines observed with the PENELLOPE program \citep{manara2021}. We fit each line/observation independently, and the results are shown in Table~\ref{tab:flow_model_results_individual} and Figure~\ref{fig:flow_fit1}-\ref{fig:flow_fit2}.
 
To calculate the final parameters for each star, we first combined the fitting results from the lines observed with the same instrument on the same epoch. This is done by combining the $\chi^2$ of a given set of parameter $i$ as
\begin{equation}
\chi^2_{\mathrm{i, total}}=\sum_{j=1}^3\frac{\chi^2_{i, j}}{\chi^2_{0, j}},
\end{equation}
 where $\chi^2_{0}$ is the minimum value in a given grid and $j=1,2,3$ refers to {\halpha}, {H$\beta$}, and {H$\gamma$}, respectively. We then take the average of the model parameters from the top 100 best fits using $L=e^{-\chi_{\mathrm{total}}^2/2}$ as the weight, and the results are shown as ``Comb'' in Table~\ref{tab:flow_model_results_individual}. Finally, we calculate the weighted average of the parameters from all instruments/epochs and report the results in Table~\ref{tab:flow_model_results}.

\startlongtable
\begin{deluxetable}{lccccCCCCC}
\tablecaption{Results of Accretion Flow Fitting for Individual Lines \label{tab:flow_model_results_individual}}
\tablehead{
\colhead{Target} 	&
\colhead{Instrument} &
\colhead{Epoch} &
\colhead{Obs.Date\tablenotemark{a}} &
\colhead{Line\tablenotemark{b}} &
\colhead{\mdot} &
\colhead{\ri} &
\colhead{\rw} &
\colhead{\tmax} &
\colhead{$i$} \\
\colhead{} 	&
\colhead{} &
\colhead{} &
\colhead{UT} &
\colhead{} &
\colhead{($10^{-8}\,\msunyr$)} &
\colhead{(R$_{\star}$)} &
\colhead{(R$_{\star}$)} &
\colhead{($10^3\,$K)} &
\colhead{(deg)}
}
\startdata
CVSO58   & UVES     & 1 & 2020-11-30 & H$\alpha$ & 2.6\pm1.2 & 3.7\pm0.9 & 1.5\pm1.0 & 8.2\pm0.8 & 48\pm20  \\ 
         &          &   & ($\Delta$MJD=-1.4) & H$\beta$  & 3.6\pm1.2 & 2.5\pm0.4 & 1.3\pm0.8 & 9.3\pm0.6 & 64\pm17  \\ 
         &          &   &            & H$\gamma$ & 3.8\pm1.0 & 3.1\pm0.8 & 0.6\pm0.6 & 9.4\pm0.6 & 69\pm9  \\ 
         &          &   &            & Comb      & 2.5\pm1.4 & 3.2\pm0.9 & 0.2\pm0.1 & 7.7\pm0.6 & 63\pm12  \\ \hline
CVSO58   & UVES     & 2 & 2020-12-01 & H$\alpha$ & 2.7\pm1.4 & 3.3\pm1.1 & 1.1\pm0.9 & 8.0\pm0.9 & 44\pm22  \\ 
         &          &   & ($\Delta$MJD=-0.4) & H$\beta$  & 3.3\pm1.2 & 2.4\pm0.5 & 1.0\pm0.6 & 9.1\pm0.8 & 55\pm21  \\ 
         &          &   &            & H$\gamma$ & 3.4\pm1.3 & 2.6\pm0.5 & 0.7\pm0.5 & 8.9\pm0.8 & 56\pm15  \\ 
         &          &   &            & Comb      & 2.7\pm1.3 & 3.3\pm0.9 & 0.2\pm0.2 & 7.5\pm0.5 & 63\pm14  \\ \hline
CVSO58   & UVES     & 3 & 2020-12-02 & H$\alpha$ & 2.5\pm1.2 & 2.6\pm0.6 & 1.6\pm0.9 & 8.7\pm0.9 & 39\pm18  \\ 
         &          &   & ($\Delta$MJD=0.6) & H$\beta$  & 3.7\pm1.1 & 2.2\pm0.3 & 1.1\pm0.6 & 9.3\pm0.7 & 51\pm21  \\ 
         &          &   &            & H$\gamma$ & 3.7\pm1.2 & 2.8\pm0.6 & 0.9\pm0.7 & 9.2\pm0.7 & 64\pm11  \\ 
         &          &   &            & Comb      & 2.6\pm1.3 & 2.6\pm0.7 & 0.3\pm0.3 & 8.0\pm0.8 & 59\pm15  \\ \hline
CVSO58   & XS       & 0 & 2020-12-02 & H$\alpha$ & 2.5\pm1.3 & 3.5\pm1.0 & 1.2\pm0.9 & 8.2\pm0.8 & 42\pm21  \\ 
         &          &   & ($\Delta$MJD=0.7) & H$\beta$  & 3.9\pm1.1 & 2.2\pm0.3 & 1.0\pm0.6 & 9.4\pm0.6 & 52\pm21  \\ 
         &          &   &            & H$\gamma$ & 3.7\pm1.2 & 2.5\pm0.5 & 0.6\pm0.5 & 9.2\pm0.8 & 67\pm11  \\ 
         &          &   &            & Comb      & 2.5\pm1.3 & 2.2\pm0.5 & 0.3\pm0.3 & 8.6\pm0.8 & 58\pm14  \\ \hline
CVSO90   & ESPRESSO & 1 & 2020-12-15 & H$\alpha$ & 2.5\pm1.4 & 4.9\pm0.8 & 0.7\pm0.7 & 7.9\pm0.7 & 37\pm9  \\ 
         &          &   & ($\Delta$MJD=-1.3) & H$\beta$  & 2.6\pm1.2 & 5.7\pm0.6 & 0.9\pm0.8 & 8.6\pm0.8 & 36\pm7  \\ 
         &          &   &            & H$\gamma$ & 3.2\pm1.1 & 4.7\pm0.8 & 0.2\pm0.0 & 8.4\pm0.5 & 44\pm10  \\ 
         &          &   &            & Comb      & 2.3\pm1.2 & 4.7\pm0.8 & 0.2\pm0.0 & 8.5\pm0.6 & 40\pm9  \\ \hline
CVSO90   & XS       & 0 & 2020-12-15 & H$\alpha$ & 2.5\pm1.4 & 5.0\pm0.8 & 0.7\pm0.7 & 7.9\pm0.7 & 37\pm9  \\ 
         &          &   & ($\Delta$MJD=-1.4) & H$\beta$  & 2.7\pm1.2 & 5.7\pm0.5 & 1.1\pm0.8 & 8.8\pm0.7 & 36\pm7  \\ 
         &          &   &            & H$\gamma$ & 3.5\pm1.1 & 4.9\pm0.8 & 0.2\pm0.0 & 8.4\pm0.3 & 45\pm11  \\ 
         &          &   &            & Comb      & 2.8\pm1.3 & 4.9\pm0.8 & 0.2\pm0.0 & 8.4\pm0.5 & 42\pm10  \\ \hline
CVSO104  & UVES     & 1 & 2020-11-25 & H$\alpha$ & 2.5\pm1.3 & 3.2\pm0.5 & 1.9\pm0.7 & 8.5\pm0.8 & 52\pm14  \\ 
         &          &   & ($\Delta$MJD=-1.4) & H$\beta$  & 3.1\pm1.1 & 4.0\pm0.5 & 1.7\pm0.8 & 9.0\pm0.6 & 59\pm10  \\ 
         &          &   &            & H$\gamma$ & 3.4\pm1.0 & 4.8\pm0.8 & 1.7\pm0.8 & 9.2\pm0.6 & 56\pm8  \\ 
         &          &   &            & Comb      & 2.0\pm1.2 & 3.3\pm1.0 & 0.2\pm0.0 & 8.4\pm0.9 & 63\pm15  \\ \hline
CVSO104  & UVES     & 2 & 2020-11-26 & H$\alpha$ & 2.7\pm1.3 & 2.4\pm0.6 & 1.2\pm0.9 & 8.0\pm0.9 & 50\pm17  \\ 
         &          &   & ($\Delta$MJD=-0.4) & H$\beta$  & 3.9\pm1.0 & 5.0\pm0.8 & 0.6\pm0.7 & 8.8\pm0.5 & 62\pm16  \\ 
         &          &   &            & Comb      & 2.7\pm1.3 & 2.4\pm0.6 & 1.2\pm1.0 & 8.0\pm0.9 & 51\pm16  \\ \hline
CVSO104  & UVES     & 3 & 2020-11-27 & H$\alpha$ & 2.6\pm1.2 & 2.6\pm0.8 & 1.0\pm0.9 & 8.0\pm0.9 & 51\pm17  \\ 
         &          &   & ($\Delta$MJD=0.5) & H$\beta$  & 3.0\pm1.2 & 2.6\pm0.3 & 1.1\pm0.7 & 9.1\pm0.7 & 41\pm14  \\ 
         &          &   &            & H$\gamma$ & 3.7\pm1.1 & 4.0\pm1.1 & 1.3\pm0.6 & 9.6\pm0.5 & 48\pm15  \\ 
         &          &   &            & Comb      & 2.1\pm1.2 & 2.5\pm0.9 & 0.3\pm0.2 & 8.8\pm0.8 & 50\pm18  \\ \hline
CVSO104  & UVES     & 4 & 2021-01-29 & H$\alpha$ & 2.6\pm1.3 & 2.3\pm0.7 & 1.1\pm0.8 & 8.0\pm0.8 & 61\pm15  \\ 
         &          &   & ($\Delta$MJD=63.5) & H$\beta$  & 3.3\pm1.2 & 2.2\pm0.6 & 0.7\pm0.5 & 8.9\pm0.9 & 55\pm18  \\ 
         &          &   &            & H$\gamma$ & 3.1\pm1.3 & 2.8\pm1.2 & 0.4\pm0.3 & 9.2\pm0.7 & 46\pm15  \\ 
         &          &   &            & Comb      & 2.5\pm1.2 & 2.7\pm1.2 & 0.2\pm0.0 & 8.6\pm1.0 & 69\pm11  \\ \hline
CVSO104  & XS       & 0 & 2020-11-27 & H$\alpha$ & 2.8\pm1.2 & 3.4\pm1.0 & 1.2\pm0.9 & 7.6\pm0.5 & 54\pm17  \\ 
         &          &   & ($\Delta$MJD=0.6) & H$\beta$  & 3.2\pm1.3 & 3.4\pm0.5 & 1.2\pm0.9 & 8.8\pm0.9 & 56\pm17  \\ 
         &          &   &            & H$\gamma$ & 2.7\pm1.3 & 4.9\pm0.9 & 0.3\pm0.3 & 8.8\pm0.6 & 52\pm16  \\ 
         &          &   &            & Comb      & 1.2\pm0.4 & 4.1\pm1.2 & 0.2\pm0.0 & 9.7\pm0.4 & 42\pm15  \\ \hline
CVSO107  & UVES     & 1 & 2020-12-03 & H$\alpha$ & 2.5\pm1.2 & 3.3\pm0.9 & 1.6\pm0.9 & 8.0\pm0.7 & 46\pm20  \\ 
         &          &   & ($\Delta$MJD=-1.3) & H$\beta$  & 3.0\pm1.3 & 2.6\pm0.5 & 2.2\pm0.5 & 9.0\pm0.8 & 56\pm16  \\ 
         &          &   &            & H$\gamma$ & 2.9\pm1.3 & 3.8\pm0.3 & 2.0\pm0.7 & 8.8\pm0.7 & 56\pm6  \\ 
         &          &   &            & Comb      & 2.4\pm1.2 & 3.1\pm0.7 & 1.3\pm0.9 & 7.9\pm0.8 & 48\pm18  \\ \hline
CVSO107  & UVES     & 2 & 2020-12-04 & H$\alpha$ & 2.5\pm1.3 & 2.6\pm0.8 & 1.6\pm0.8 & 8.0\pm0.7 & 38\pm18  \\ 
         &          &   & ($\Delta$MJD=-0.4) & H$\beta$  & 3.2\pm1.1 & 2.3\pm0.4 & 1.2\pm0.8 & 8.8\pm0.9 & 44\pm19  \\ 
         &          &   &            & H$\gamma$ & 3.5\pm1.2 & 2.7\pm0.6 & 0.8\pm0.6 & 9.1\pm0.8 & 41\pm17  \\ 
         &          &   &            & Comb      & 2.3\pm1.0 & 2.5\pm0.6 & 0.4\pm0.6 & 7.3\pm0.5 & 54\pm15  \\ \hline
CVSO107  & UVES     & 3 & 2020-12-05 & H$\alpha$ & 2.5\pm1.2 & 2.9\pm1.1 & 1.5\pm0.9 & 7.9\pm0.7 & 36\pm16  \\ 
         &          &   & ($\Delta$MJD=0.7) & H$\beta$  & 3.1\pm1.2 & 2.8\pm0.4 & 1.5\pm0.8 & 8.3\pm0.6 & 54\pm19  \\ 
         &          &   &            & H$\gamma$ & 3.3\pm1.2 & 3.5\pm0.5 & 1.2\pm0.9 & 8.6\pm0.9 & 54\pm16  \\ 
         &          &   &            & Comb      & 2.1\pm0.9 & 3.4\pm1.0 & 0.4\pm0.6 & 7.2\pm0.4 & 52\pm20  \\ \hline
CVSO107  & XS       & 0 & 2020-12-04 & H$\alpha$ & 2.6\pm1.3 & 2.6\pm0.8 & 1.6\pm0.8 & 8.0\pm0.7 & 38\pm19  \\ 
         &          &   & ($\Delta$MJD=-0.4) & H$\beta$  & 3.4\pm1.1 & 2.1\pm0.3 & 1.1\pm0.7 & 9.1\pm0.9 & 45\pm18  \\ 
         &          &   &            & H$\gamma$ & 3.2\pm1.4 & 2.8\pm0.7 & 0.8\pm0.7 & 9.0\pm0.9 & 38\pm19  \\ 
         &          &   &            & Comb      & 2.2\pm1.0 & 2.5\pm0.5 & 0.2\pm0.2 & 7.1\pm0.3 & 57\pm12  \\ \hline
CVSO146  & ESPRESSO & 1 & 2020-12-09 & H$\alpha$ & 0.6\pm0.3 & 2.9\pm0.6 & 1.8\pm0.7 & 9.8\pm0.9 & 37\pm14  \\ 
         &          &   & ($\Delta$MJD=-1.4) & H$\beta$  & 0.7\pm0.3 & 2.2\pm0.4 & 0.3\pm0.3 & 9.5\pm0.9 & 70\pm7  \\ 
         &          &   &            & H$\gamma$ & 0.8\pm0.2 & 2.1\pm0.3 & 0.2\pm0.0 & 9.7\pm0.8 & 57\pm13  \\ 
         &          &   &            & Comb      & 0.5\pm0.3 & 2.6\pm0.8 & 0.3\pm0.3 & 9.2\pm1.2 & 56\pm14  \\ \hline
CVSO146  & ESPRESSO & 2 & 2020-12-10 & H$\alpha$ & 0.6\pm0.3 & 3.0\pm0.6 & 1.6\pm0.8 & 9.6\pm0.9 & 54\pm12  \\ 
         &          &   & ($\Delta$MJD=-0.4) & H$\beta$  & 0.8\pm0.2 & 2.7\pm0.3 & 1.3\pm0.7 & 10.0\pm0.7 & 72\pm6  \\ 
         &          &   &            & H$\gamma$ & 0.8\pm0.2 & 2.1\pm0.3 & 0.2\pm0.0 & 9.8\pm0.8 & 60\pm10  \\ 
         &          &   &            & Comb      & 0.5\pm0.3 & 2.9\pm0.7 & 0.4\pm0.3 & 9.2\pm1.2 & 65\pm9  \\ \hline
CVSO146  & ESPRESSO & 3 & 2020-12-11 & H$\alpha$ & 0.6\pm0.2 & 3.3\pm0.8 & 1.9\pm0.7 & 9.9\pm0.7 & 58\pm5  \\ 
         &          &   & ($\Delta$MJD=0.6) & H$\beta$  & 0.9\pm0.2 & 2.8\pm0.0 & 1.5\pm0.6 & 10.4\pm0.5 & 75\pm0  \\ 
         &          &   &            & H$\gamma$ & 0.8\pm0.2 & 3.2\pm0.5 & 0.8\pm0.5 & 10.1\pm0.7 & 72\pm6  \\ 
         &          &   &            & Comb      & 0.5\pm0.2 & 3.2\pm0.6 & 0.7\pm0.5 & 9.6\pm1.0 & 69\pm7  \\ \hline
CVSO146  & XS       & 0 & 2020-12-09 & H$\alpha$ & 0.7\pm0.3 & 2.9\pm0.8 & 1.9\pm0.7 & 9.8\pm0.9 & 41\pm14  \\ 
         &          &   & ($\Delta$MJD=-1.4) & H$\beta$  & 0.7\pm0.2 & 2.3\pm0.4 & 0.6\pm0.4 & 9.6\pm0.9 & 69\pm8  \\ 
         &          &   &            & H$\gamma$ & 0.8\pm0.2 & 2.1\pm0.3 & 0.2\pm0.0 & 9.7\pm0.9 & 59\pm13  \\ 
         &          &   &            & Comb      & 0.4\pm0.2 & 2.6\pm0.8 & 0.2\pm0.1 & 9.1\pm1.2 & 59\pm13  \\ \hline
CVSO176  & UVES     & 1 & 2020-11-28 & H$\alpha$ & 3.5\pm2.1 & 3.4\pm0.8 & 1.4\pm0.8 & 7.8\pm0.7 & 38\pm20  \\ 
         &          &   & ($\Delta$MJD=-0.7) & H$\beta$  & 4.1\pm2.1 & 3.3\pm0.4 & 1.7\pm0.7 & 8.6\pm0.8 & 59\pm14  \\ 
         &          &   &            & H$\gamma$ & 4.6\pm1.8 & 4.9\pm0.7 & 1.8\pm0.7 & 8.8\pm0.7 & 60\pm7  \\ 
         &          &   &            & Comb      & 3.0\pm2.1 & 2.2\pm0.5 & 0.2\pm0.0 & 8.0\pm0.7 & 42\pm21  \\ \hline
CVSO176  & UVES     & 2 & 2020-11-29 & H$\alpha$ & 3.4\pm2.2 & 3.9\pm1.0 & 1.4\pm0.9 & 7.5\pm0.5 & 45\pm19  \\ 
         &          &   & ($\Delta$MJD=0.2) & H$\beta$  & 4.1\pm2.2 & 3.3\pm0.7 & 2.0\pm0.7 & 8.5\pm0.7 & 63\pm8  \\ 
         &          &   &            & H$\gamma$ & 3.8\pm1.9 & 4.8\pm0.8 & 1.9\pm0.8 & 8.9\pm0.7 & 56\pm8  \\ 
         &          &   &            & Comb      & 3.0\pm2.1 & 2.9\pm1.1 & 0.2\pm0.2 & 7.3\pm0.7 & 55\pm20  \\ \hline
CVSO176  & UVES     & 3 & 2020-11-30 & H$\alpha$ & 3.6\pm2.2 & 3.6\pm0.9 & 1.4\pm0.8 & 7.6\pm0.5 & 42\pm19  \\ 
         &          &   & ($\Delta$MJD=1.3) & H$\beta$  & 4.0\pm1.9 & 2.8\pm0.2 & 1.8\pm0.8 & 8.9\pm0.8 & 61\pm13  \\ 
         &          &   &            & H$\gamma$ & 5.0\pm1.8 & 4.3\pm0.8 & 1.6\pm0.8 & 9.0\pm0.6 & 70\pm7  \\ 
         &          &   &            & Comb      & 2.7\pm1.7 & 2.7\pm0.7 & 0.2\pm0.1 & 7.5\pm0.7 & 57\pm20  \\ \hline
CVSO176  & XS       & 0 & 2020-12-02 & H$\alpha$ & 3.8\pm2.3 & 3.4\pm0.8 & 1.5\pm0.8 & 7.7\pm0.6 & 39\pm20  \\ 
         &          &   & ($\Delta$MJD=3.3) & H$\beta$  & 5.5\pm1.6 & 2.0\pm0.0 & 2.1\pm0.6 & 9.4\pm0.5 & 41\pm20  \\ 
         &          &   &            & H$\gamma$ & 5.4\pm1.8 & 4.7\pm1.3 & 1.0\pm0.5 & 9.3\pm0.6 & 37\pm20  \\ 
         &          &   &            & Comb      & 2.9\pm1.7 & 3.9\pm1.6 & 0.3\pm0.3 & 9.6\pm0.6 & 59\pm22 

\enddata
\tablenotetext{a}{The numbers in parenthesis are the difference (in days) between the VLT observations and the COS observations, $\Delta\mathrm{MJD}=\mathrm{MJD}_{\mathrm{VLT}}-\mathrm{MJD}_{\mathrm{COS}}$.}
\tablenotetext{b}{``Comb'' refers to the fitting results where the lines in the same epoch are fitted simultaneously using the summed $\chi^2$ as described in the text.}
\end{deluxetable}

\end{document}